\documentclass[a4paper,11pt]{article}
\pdfoutput=1
\usepackage{jcappub}
\usepackage[T1]{fontenc}
\usepackage{booktabs}

\title{Hybrid radio and particle detection of air showers: potential for ultra-high-energy photon identification}

\author[a,b,1]{Paul Minodier,\note{Corresponding author.}}
\author[c]{Kaoru Takahashi,}
\author[a]{Kumiko Kotera}
\author[c]{and Takashi Sako}

\affiliation[a]{Institut d’Astrophysique de Paris, CNRS UMR 7095, Sorbonne Université, 98bis bd Arago 75014, Paris,
France}
\affiliation[b]{ILANCE, CNRS – University of Tokyo International Research Laboratory, Kashiwa, Chiba 277-8582, Japan}
\affiliation[c]{Institute for Cosmic Ray Research, University of Tokyo, Kashiwa 277-8582, Japan}

\emailAdd{minodier@iap.fr}

\abstract{The autonomous radio-detection of extensive air showers initiated by ultra-high-energy (UHE) particles arriving with very inclined zenith angles has seen significant advancements in recent years, with several large-scale surface arrays planned and prototypes already in operation. In this work, we examine whether these radio detectors, supplemented by scintillators, could serve as competitive UHE photon detectors. Indeed, for inclined showers, radio emissions can be detected by antennas for both cosmic-ray and photon primaries, while the muon-rich signatures of the former would typically trigger the scintillators. Using two key observables — the total root mean square of the radio signal and the total energy deposit recorded in the scintillators — we show that effective separation between the two types of showers could be achieved in a hybrid radio antenna and scintillator setup. As a case study, we apply this method to a hypothetical hybrid array of radio antennas, complemented by Telescope Array-like scintillators, with the layout of the prototype of the Giant Radio Array for Neutrino Detection (GRAND), GRANDProto300. Our estimates show that such a hybrid array could set competitive upper limits on the integral photon flux in the energy range of approximately 0.3 to 3 EeV.}

\begin{document}
\maketitle
\flushbottom

\section{Introduction}

Ultra-high-energy (UHE) photons, with energies above $10^{17}\,\textrm{eV}$, are one of the long-sought messengers to pinpoint and probe the sources of UHE energy cosmic rays \cite{2019FrASS...6...23B,Alves_2019,PhysRevLett.123.051101,2021Natur.594...33C,PhysRevD.100.103008,annurev-nucl-112822-025357,Cao_2024}.
They can be produced by UHE cosmic rays via photo-hadronic or hadronic interactions either on the radiative or baryonic backgrounds of the source environment, or in the course of their propagation to the Earth, by interaction on cosmic background photons. Above $10^{17}\,\textrm{eV}$, interactions with the Extragalactic Background Light through pair-production processes limit the UHE photon horizon to roughly the distance of the Local Group galaxies \cite{PhysRevD.58.043004,doi:10.1142/S0217732307022864}. Therefore, the detection of these photons would evidence the presence of one or several UHE accelerators within the Local Group. Setting limits on non-detection can also provide important constraints on the Galactic or extragalactic origin of cosmic rays with energy $10^{17-18.5}\,\textrm{eV}$. 

The LHAASO experiment recently detected photons with energies up to $2.5 \, \textrm{PeV}$, in the direction of the star-forming region Cygnus X \cite{LHAASOCOLLABORATION2024449}, providing strong evidence of cosmic-ray acceleration up to energies above $10^{15}\,\textrm{eV}$ in the Galaxy.
No significant detection has been reported at higher energies, although several experiments have  set constraints on the photon flux above $10^{17}\,\textrm{eV}$, including KASCADE-Grande \cite{Apel_2017} and Telescope Array \cite{ABBASI20198, Abbasi:2021Z9, PhysRevD.88.112005, 2025arXiv251201638T} in the Northern Hemisphere, and the Pierre Auger Observatory \cite{AbdulHalim_2025,Abreu_2023,Abreu_2022,PhysRevD.110.062005, AbdulHalim:2023SN,Savina:2021mS,Aab_2017} in the Southern Hemisphere.

UHE particles can be detected via the extensive air-showers of secondary particles they create upon entering the atmosphere.
Because of the dominant cosmic-ray background, searches for UHE photons require powerful discrimination methods, based on the differences between photon- and proton-induced showers. The number of muons in the shower is a key indicator for the primary type \cite{GONZALEZ202048}: while photon-induced showers develop mostly through electromagnetic interaction channels, proton-induced showers exhibit important hadronic components, with charged pions and kaons decaying into muons, leading to a larger number of muons. Other features, like the atmospheric depth of maximum shower development $X_{\rm max}$ or the lateral spread of the particle profiles, have also been used in previous studies. 

In the last decade, radio-detection of UHE cosmic showers has drastically improved \cite{SCHRODER20171,HUEGE20161}, with the successful deployment and operation of large-scale arrays such as Auger-Radio \cite{Aab_2018, AbdulHalim:2025Zz, AbdulHalim:2023ZX} and GRAND~\cite{2025arXiv250921306G,2025arXiv250709585A}. The radio technique appears natural to efficiently detect and probe the electromagnetic content of photon-generated showers, for which particle trigger is challenging. 

Interestingly, these new generation experiments focus on the detection of very inclined showers, arriving with zenith angle above $60^\circ$. 
Because of their longer path through the atmosphere, the electromagnetic and hadronic components of very inclined showers are absorbed before reaching the ground \cite{Holt2018_1000083318}. Only muon-rich showers, mostly generated by cosmic-rays, can be detected by particle detectors on the ground. Hence these can act as vetoes to discriminate between cosmic rays and photons upon radio-detection of a shower. The muon content of a shower is a powerful indicator of the primary type, already used in previous photon searches \cite{AbdulHalim_2025, PhysRevD.110.062005, PhysRevLett.123.051101, 2021ExA....52...85K}.

Hybrid detection using radio and particle detectors has already been investigated at the Pierre Auger Observatory for mass estimation of very inclined showers \cite{Holt2018_1000083318,2q9f-pbrp}.
Because radio antennas and surface detectors like scintillation detectors or water-Cherenkov detectors are relatively cheap and easy to install compared to underground muon detectors, this option is scalable and could be used for large-scale arrays.

In this paper, we present a method to distinguish between photon- and proton-induced showers using a hybrid array of radio antennas and scintillation detectors, by combining the radio signal measured by the antennas with the particle energy deposit in scintillators. As a case study, we apply this method to a hypothetical hybrid array between the Giant Radio Array for Neutrino Detection (GRAND) prototype, GRANDProto300, and Telescope Array. The hybrid detection of proton and photon showers using radio antennas and particle detectors is studied in section \ref{sec:radio}. The discrimination method is presented in section \ref{sec:discr}, along with its performances and the associated upper limits on the photon flux in the case of a hybrid array of radio antennas and scintillation detectors with the layout of GRANDProto300.

\section{Radio and particle detection of very inclined photon and proton showers}\label{sec:radio}

In this section, we present physical signals specific to proton- and photon-induced air-showers, for radio and particle detection. We examine how these characteristics can be exploited in the framework of a virtual experimental setup, combining two existing instrument models. We describe this illustrative setup and the simulation sets used to evaluate the triggering efficiency of UHE photons on a hybrid radio and particle detector array.

\subsection{Proton- and photon-induced inclined shower emission characteristics in radio and particles}\label{sec:signal}

The main contribution to the radio emission from air showers stems from \textit{geomagnetic} emission. It is produced by electrons and positrons drifting laterally at the shower front, under the competing influences of magnetic deflection by the Earth's magnetic field and particle scattering
off air molecules \cite{SCHOLTEN200894}. It generates a time-varying transverse current \cite{SCHOLTEN200894}, resulting in radio emission. The main contribution stems from electrons and positrons because of their higher charge/mass ratio. 

In the case of very inclined showers, as
recently demonstrated in Ref.~\cite{PhysRevLett.132.231001},
two additional mechanisms come into play. Particles experience stronger magnetic deflection because air showers develop in the less dense upper atmosphere, where Coulomb scattering is reduced and mean free paths are longer, allowing particles to propagate over greater distances.
This results first in electrons and positrons undergoing fractions of helical trajectories to
emit \textit{geo-synchrotron} radiation, with a contribution starting from frequencies of tens of $\textrm{MHz}$ for showers with a zenith angle $\theta > 68^\circ$ \cite{PhysRevLett.132.231001}. This synchrotron emission can account for $\sim10\%$ for very inclined showers. Note that the traditional \textit{Askaryan} (or \textit{charge-excess}) emission, which results from the accumulation of ionizing electrons in the shower front, represents less than $5\%$ of the total radio emission for very inclined showers \cite{PhysRevD.89.052002,CHICHE2022102696,Schluter_2023}. 
Second, the stronger magnetic deflections lead to larger shower lateral extent. When the lateral extent exceeds the coherence length of the radio signal, the emission becomes incoherent and the strength of the radio signal drops at low frequencies \cite{PhysRevLett.132.231001,Guelfand_2024}. 

Because the emitting particles are relativistic, this emission is strongly beamed forward, and Cherenkov effects produce a signal boosted along a cone with a typical aperture of $\sim 1^\circ$.  The radio footprint measured by radio antennas on the ground is the projection of a cone, and has a characteristic ring shape. This characteristic shape can be exploited to perform reconstruction of both the arrival direction and the energy of the primary particle, using a phenomenological model of the angular distribution function \cite{Guelfand:2025lY,GUELFAND2025103120}.

\begin{figure}
    \centering
    \includegraphics[width=\linewidth]{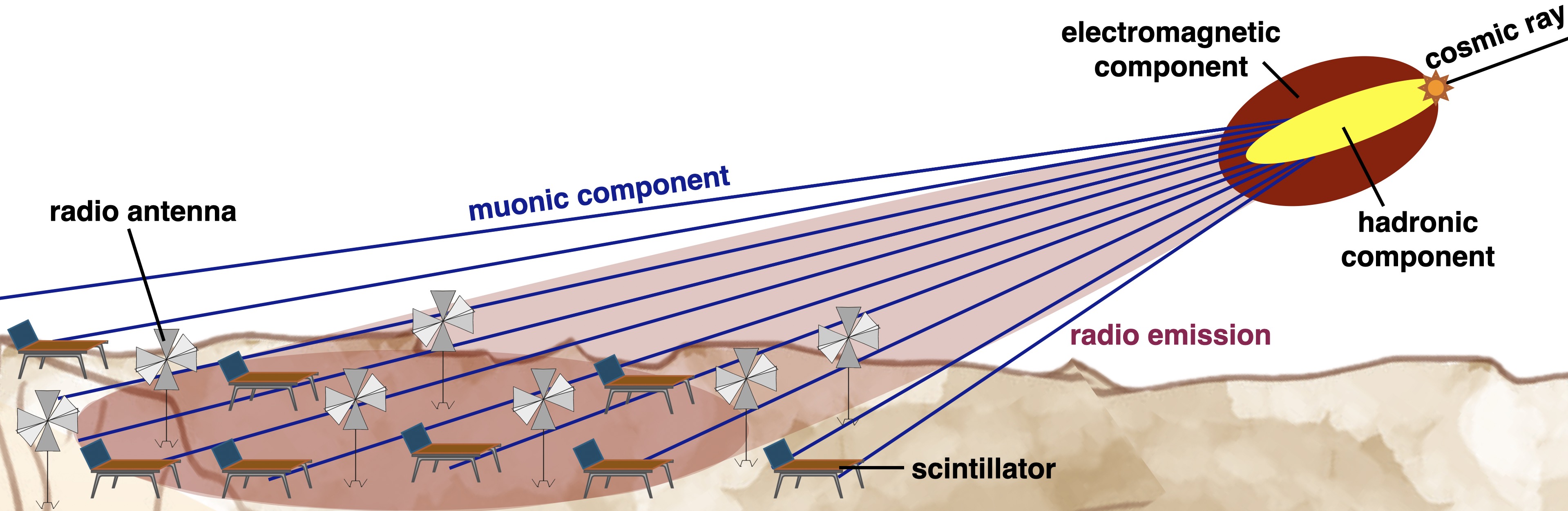}
    \caption{Main components of a very inclined air shower. Only muons and the radio emission from the electromagnetic component reach the ground and can be detected with scintillation detectors and radio antennas.}
    \label{fig:inclined_showers}
\end{figure}

For very inclined air showers, the electromagnetic and hadronic components are absorbed before reaching the ground because of the increased atmospheric depth crossed -- for a shower with zenith angle $\theta = 80^\circ$, the atmospheric depth crossed is of order $5000 \,\textrm{g}/\textrm{cm}^2$. Particles reaching the ground are mostly muons and secondary photons, few electrons, and almost no hadrons. Muons carry the bulk of the energy of particles reaching the ground, and represent the dominant contribution to the energy deposit in surface detectors on the ground \cite{Holt2018_1000083318}, as represented schematically in Figure~\ref{fig:inclined_showers}.

\begin{figure}
    \centering
    \includegraphics[width=\linewidth]{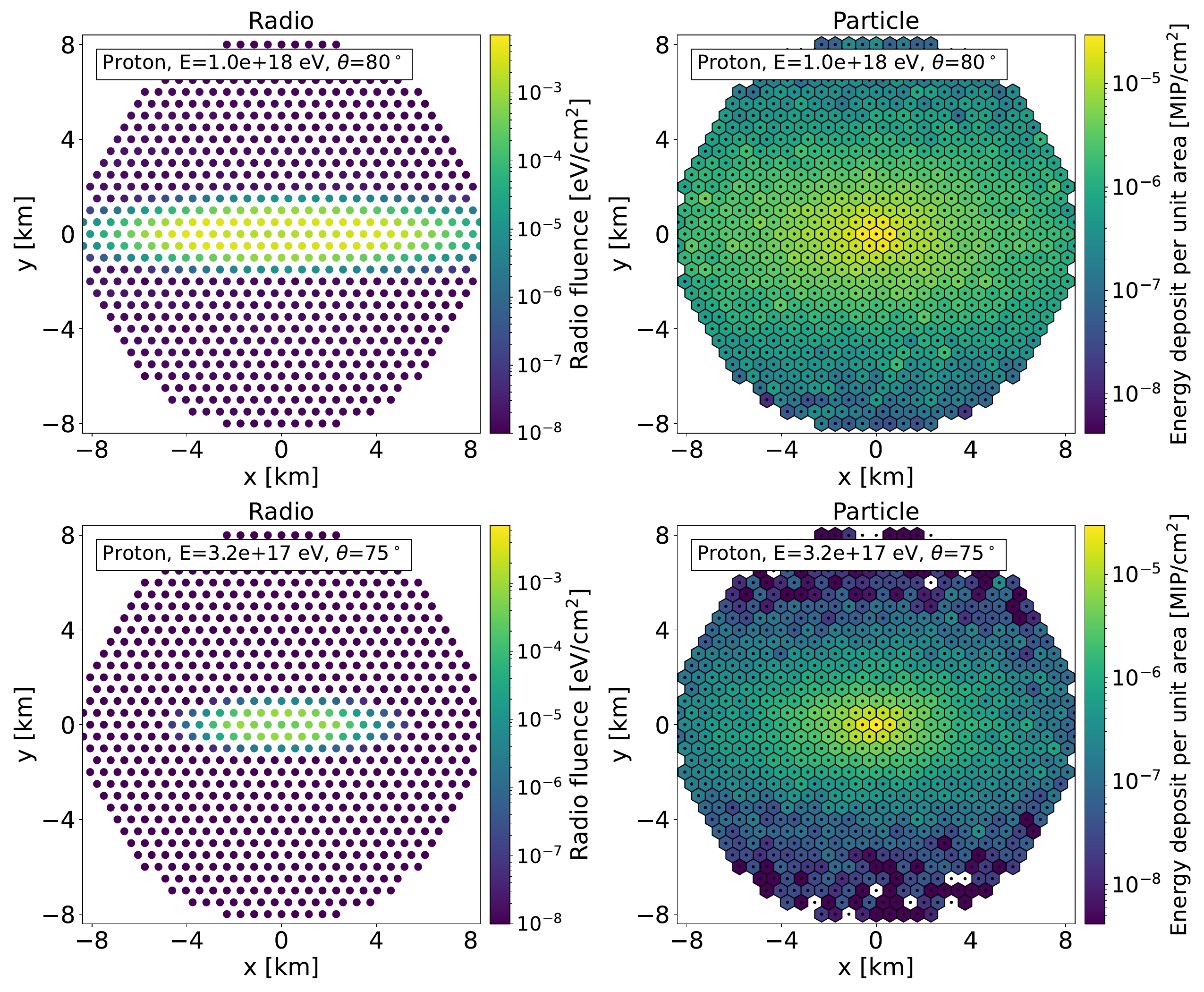}
    \caption{Examples of radio (\textit{left}) and particle (\textit{right}) footprints for two proton events, with energy $E = 10^{18} \textrm{ eV}$ and zenith $\theta = 80^{\circ}$ ({\it upper row}), and $E = 10^{17.5} \textrm{ eV}$ and $\theta = 75^{\circ}$ ({\it lower row}). The radio footprint is represented by the fluence of the electric field at antenna positions along an hexagonal grid of spacing $a=577\,$m. The particle footprint shows the average energy deposit in scintillation detectors in hexagonal bins centered around the antenna positions.}
    \label{fig:proton_events}
\end{figure}

\begin{figure}
    \centering
    \includegraphics[width=\linewidth]{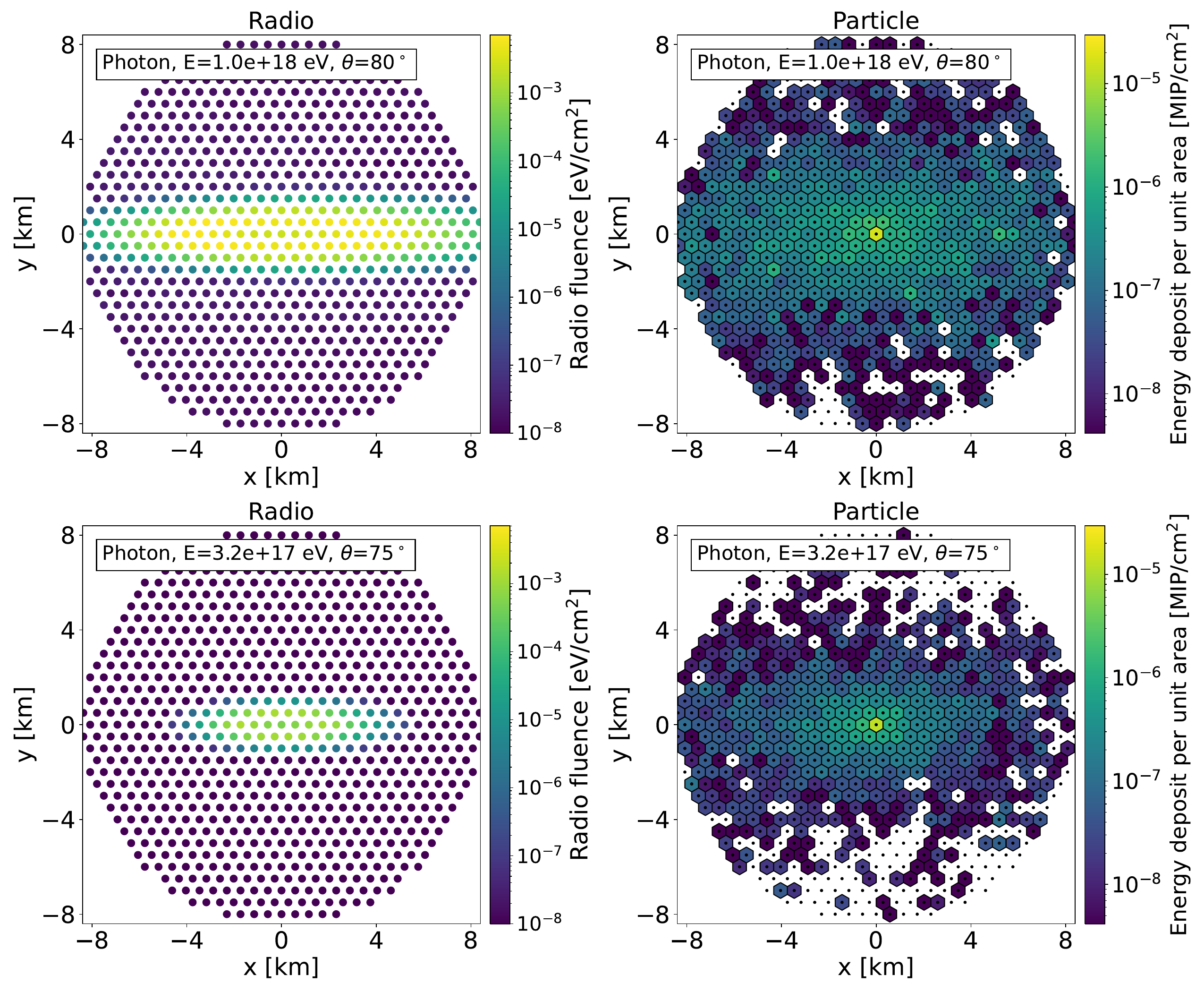}
    \caption{Same as Fig.~\ref{fig:proton_events} but for two photon events, with energy $E = 10^{18} \textrm{ eV}$ and zenith $\theta = 80^{\circ}$ ({\it upper row}), and $E = 10^{17.5} \textrm{ eV}$ and $\theta = 75^{\circ}$ ({\it lower row}).}
    \label{fig:gamma_events}
\end{figure}

As an illustration, Figures~\ref{fig:proton_events} and \ref{fig:gamma_events} show the radio and particle footprints for two proton events (Fig.~\ref{fig:proton_events}) and two photon events (Fig.~\ref{fig:gamma_events}) with energy $E = 10^{18}\,\textrm{eV}$ and zenith angle $\theta = 80^{\circ}$ (upper row), and with $E = 10^{17.5}\,\textrm{eV}$ and $\theta = 75^{\circ}$ (bottom).
The radio fluence is represented by a color-coding of dots located at antenna positions along a fictive hexagonal grid with side length of $a = 577\,$m in a circle of radius $8.4\,\rm km$. The particle footprint is obtained by computing the energy deposits in ``virtual'' scintillation detectors, with a surface area $S = 3\,\rm m^2$, arranged on a regular $6\rm\,m\times 6\,m$ grid. The energy deposit per unit area is then computed by summing the energy deposits in hexagonal bins of area $A = \sqrt{3}a^2/2 = 0.29\,\rm km^2$, and finally dividing by the total surface $S_{\rm tot}$ covered by detectors in each hexagonal bin, $S_{\rm tot} = N_{\rm scint}S$ with $N_{\rm scint} \sim A/36\,\rm m^2 $ the number of detectors in each hexagonal bin.
The energy deposit per unit area is at least 10 times lower for photon primaries than for proton ones. Additionally, because photon-initiated showers develop mostly through electromagnetic channels, their radio emission is slightly stronger than for proton primaries. The influence of the primary arrival direction is also apparent. 

\begin{figure}
    \centering
    \includegraphics[width=\linewidth]{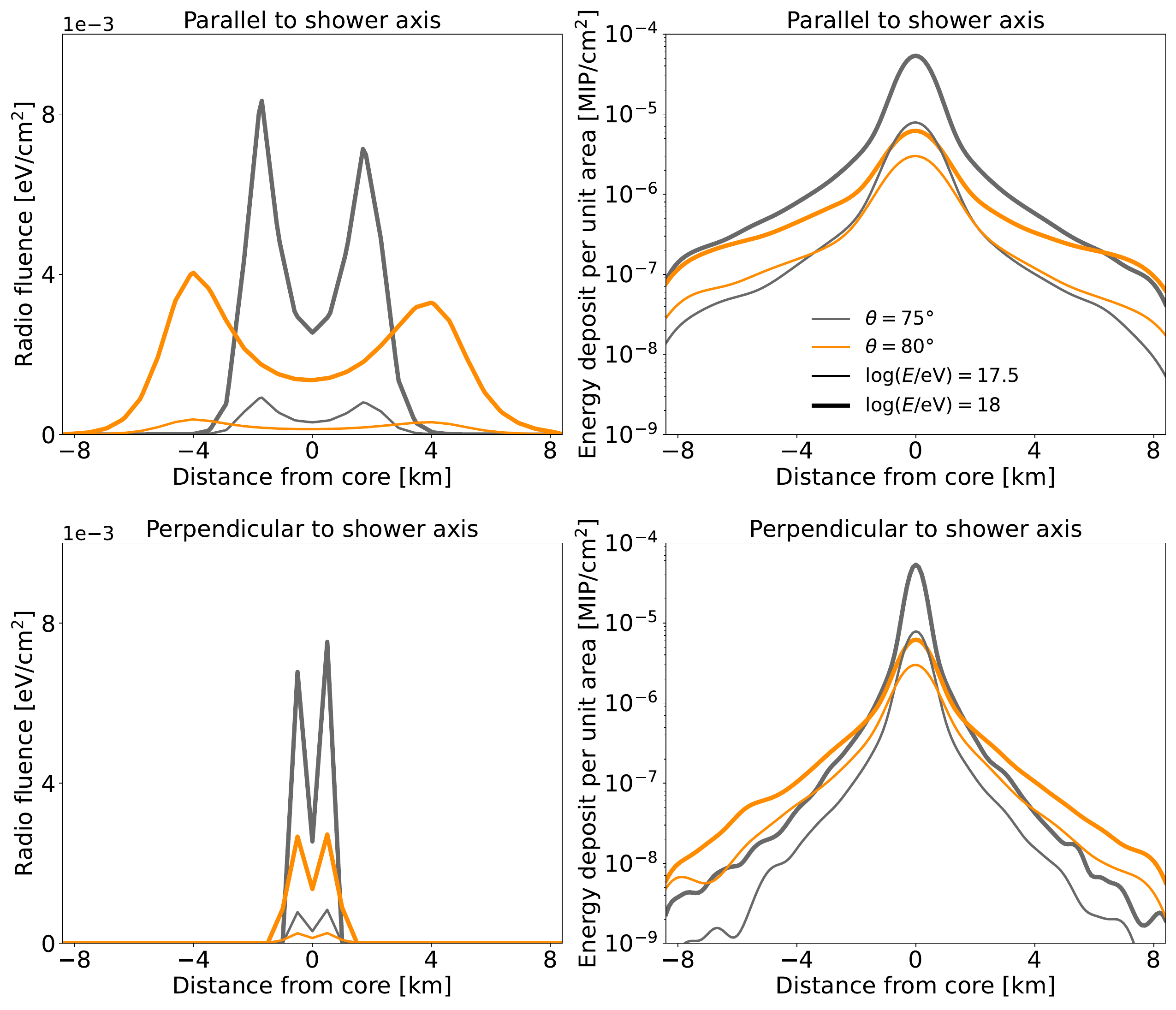}
    \caption{Radio (\textit{left}) and particle (\textit{right}) profiles, parallel (\textit{upper row}) and perpendicular (\textit{lower row}) to the shower axis, for proton events with energy $E = 10^{17.5}\textrm{ eV, }10^{18} \textrm{ eV}$ and zenith angles $\theta = 75^{\circ},80^{\circ}$.}
    \label{fig:profiles_notscaled}
\end{figure}

At a constant primary energy, inclined showers have larger radio footprints on the ground, while the particle footprint is spread out over a larger area, resulting in a lower maximum energy deposit but a more moderate decrease of the particle energy deposit away from the shower core.
These effects are further illustrated for the proton case in
Figure~\ref{fig:profiles_notscaled}, where we present the particle energy deposit and radio emission fluence profiles, both perpendicular and parallel to the shower axis, for various primary energies and zenith angles. The size of the radio footprint, i.e., the length of its major axis, roughly scales as $L\sim 2\textrm{ km}/\cos\theta$ and can reach $\sim 30\,\textrm{km}$ for very inclined showers with zenith angles close to $85^\circ$ \cite{Aab_2018}. The maximum amplitude of the radio signal is lower for more inclined showers.
For less inclined showers, the particle energy deposit is more peaked around the shower core and falls more steeply.

Proton and photon showers exhibit similar signatures in radio. Showers initiated by heavier nuclei develop earlier in the atmosphere than proton ones, while photon showers develop deeper on average due to the small multiplicity of electromagnetic interactions \cite{PhysRevD.110.062005}, and the LPM effect above $10^{18}\,$eV \cite{Landau:1953um,PhysRev.103.1811,PhysRevD.59.113012}. As a result, the difference in radio signature between heavy nuclei and photon showers will be more significant than between proton and photon showers. Moreover, heavier nuclei initiate showers with a richer muonic component. Indeed, the number of muons in the shower scales roughly as $A^{0.1\sim0.15}$, where $A$ is the mass number of the primary \cite{MATTHEWS2005387}. Therefore, proton-initiated showers are the most photon-like ones and will be the dominant source of background. When evaluating the performances of our discrimination method described in Section~\ref{sec:discr}, we will consider the conservative scenario of a full proton background.

\subsection{Hybrid experimental setup} \label{sec:setup}

To provide a quantitative illustration of the detection principles examined in this work, we base our experimental parameters on the 300-antenna prototype of GRAND, GRANDProto300 \cite{2020SCPMA..6319501A,2025arXiv250921306G,2025arXiv250709585A}, and on scintillator models of Telescope Array \cite{ABUZAYYAD201287,Udo:2017ja}.

\begin{figure}
    \centering
    \includegraphics[width=0.7\linewidth]{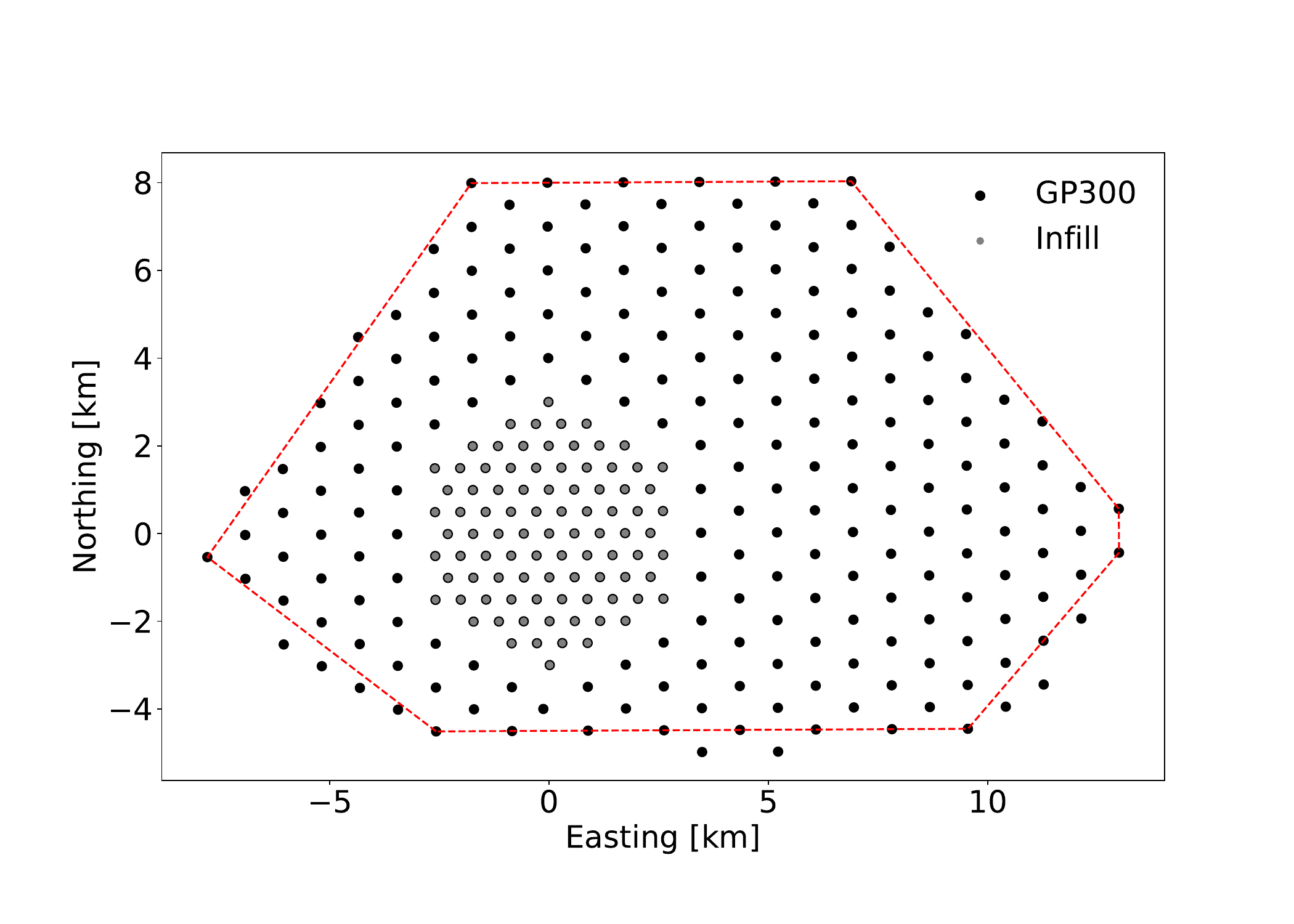}
    \caption{Detector layout of GRANDProto300, shown as black dots. Antennas in the infill array are also colored in grey. The dashed red contour marks the area in which shower core positions are randomly distributed (see Section~\ref{sec:sims}).}
    \label{fig:layout}
\end{figure}

The GRANDProto300 array is located near Dunhuang, in the Gobi desert in China, at an altitude $\sim 1200\,\textrm{m}$ above sea level. In its final configuration, it will be composed of (see Fig.~\ref{fig:layout}) an overall array of $245$ antennas with a spacing of $1\,\textrm{km}$, called the \textit{coarse} array in this study, and of a denser hexagonal array of $91$ antennas with a spacing of $577\,\textrm{m}$, called the \textit{infill}, for a total of $299$ antennas ($37$ antennas are both in the infill and the coarse arrays).
The infill array is optimized for lower energy events, for which the radio footprint will be smaller, while the coarse setup targets more inclined events. The radio antenna model of GRANDProto300 measures the electric field along three directions (East-West, North-South and vertical) in the $50-200\,\textrm{MHz}$ frequency range.

To study the performance of hybrid detection by complementing radio antennas with surface particle detectors, we take as an example Telescope Array-like scintillation detectors (SDs) \cite{ABUZAYYAD201287}. Each scintillation detector is made of two layers of plastic scintillator with a $3\,\textrm{m}^2$ area and $1.2\,\textrm{cm}$ thickness, measuring the energy deposit in units of minimum ionizing particles (MIPs). 
The hybrid array we study is obtained by considering an additional scintillation detector at each position of GRANDProto300 antennas.

In order to quantify the detection performance of this hybrid array using simulations, we consider the following trigger criteria. At detector level, an antenna is triggered if the peak-to-peak amplitude of the signal trace, filtered to keep the $50-200\,\textrm{MHz}$ frequency range, along one of the three antenna directions, is above $E_{\rm thr} = 75\: \mu\rm V/m$, i.e., roughly three times the Galactic noise in the observed frequency range. A SD detector is triggered if the total energy deposit, averaged between the upper and lower layers, is above $\epsilon_{\rm thr} = 3\textrm{ MIP}$ equivalents. At event level, an event is triggered in radio if more than $N_{\rm ant} = 4$ antennas are triggered, and in particle if more than $N_{\rm SD} = 3$ SDs are triggered.

\subsection{Radio and particle air-shower simulation sets}\label{sec:sims}

The simulations sets used in this work were computed with CORSIKA v7.7550 \cite{Heck:1998vt} and CoREAS v1.4 \cite{10.1063/1.4807534} using QGSJETII-04 \cite{PhysRevD.83.014018} for the high energy hadronic interaction model and URQMD 1.3cr \cite{BASS1998255,MBleicher_1999} for the low energies. A total of 8972 photon-induced and 8972 proton-induced events were generated uniformly in both energy and zenith angle, between $10^{17} \text{  }\mathrm{eV}$ and $10^{18.5} \text{  }\mathrm{eV}$, and $70^\circ$ and $85^\circ$. 
Azimuth angles are drawn uniformly between $0^\circ$ and $360^\circ$. Shower core positions are randomly distributed in the GRANDProto300 layout, inside the red contour shown in Figure~\ref{fig:layout}. We use the thinning option available in CORSIKA to reduce computation time by reducing the number of low-energy particles to follow, with a thinning parameter $\epsilon = 10^{-6}$ for both proton and photon showers \cite{KOBAL2001259,Ellwanger:20258S}. We simulate events using the magnetic field and atmospheric profile for Dunhuang, China. The output of the simulations is the electric field as a function of time for each antenna (with a resolution of $0.5\,\textrm{ns}$ over a window of $1.2 \,\mu \textrm{s}$, i.e., $N=2400$ time steps), projected along 3 orthonormal directions $x,y,z$, as well as all the particles reaching the observation plane, with their particle type, their position in the observation plane, their momentum, the time and their weight (because of thinning).

Each simulation is analyzed using the following procedure. For each antenna, radio traces, along the 3 directions $x, y, z$, are filtered to keep only the $50-200$ MHz frequency range. We compute the peak-to-peak amplitude of each trace (for the trigger), as well as the root mean square (RMS) of the traces: 
    \begin{equation}
        \textrm{RMS}_{j=x,y,z} = \bigg[\frac{1}{N}\sum_{i=1}^N E_j^2(t_i)\bigg]^{\frac{1}{2}} \ .
    \end{equation}
For the particle energy deposit computation, the analysis follows the TA procedure. First, a dethinning algorithm is applied to regenerate individual particles from weighted thinned particles, in order to recover small-scale structures of the shower fluctuations \cite{STOKES2012759}. The detector response is computed using look-up tables simulated with GEANT4 \cite{AGOSTINELLI2003250}, on a regular grid of virtual scintillation detectors, located every $6\,$m in a circle of radius $8.4\,\textrm{km}$ around the shower core. For each real detector in the GRANDProto300 layout, the energy deposit (in both layers) is computed by averaging the deposits of the 4 closest virtual detectors. 

\subsection{Triggering performances}\label{sec:trigger}

\begin{figure}
    \centering
    \includegraphics[width=\linewidth]{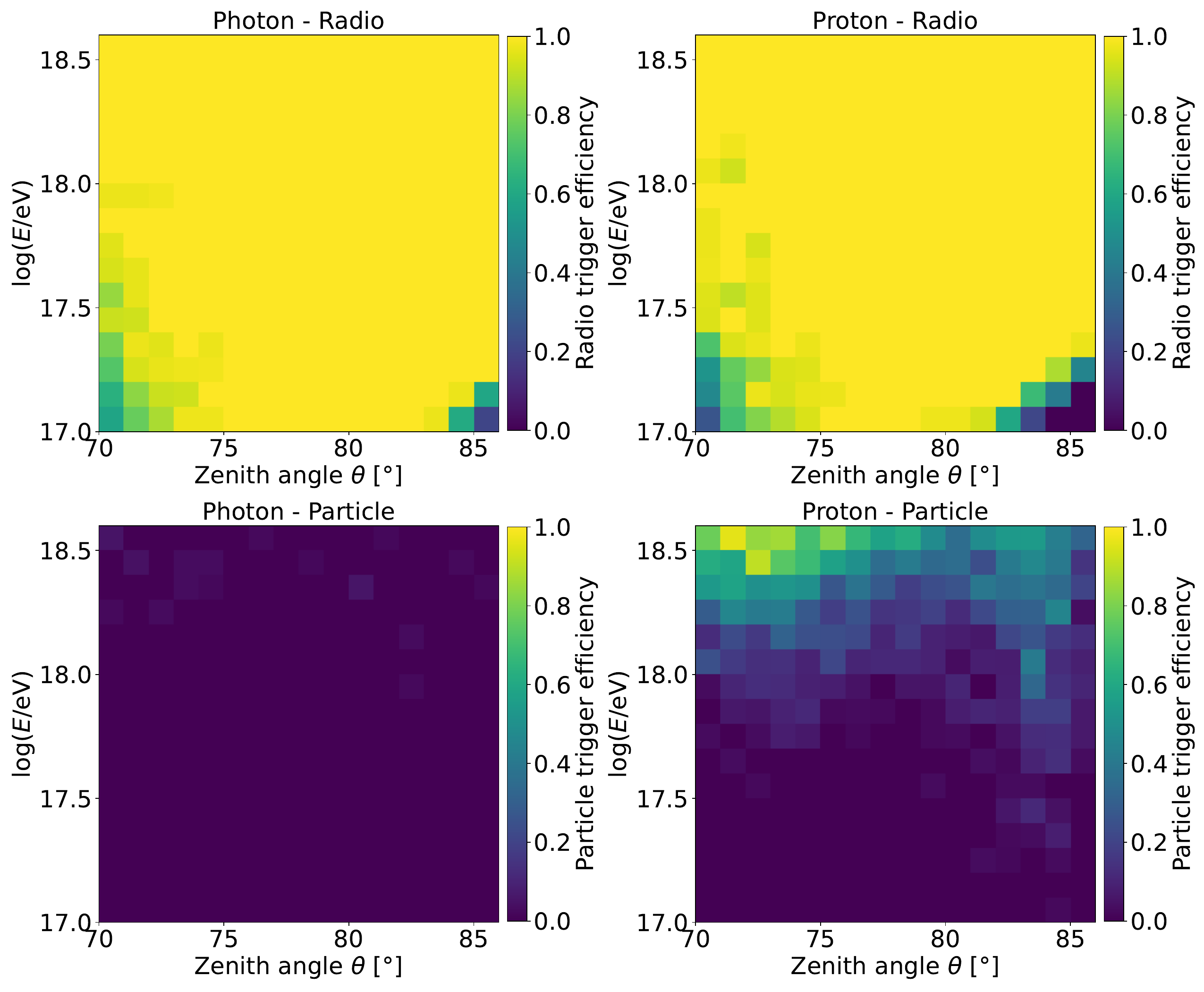}
    \caption{Trigger efficiencies for radio $r_{\rm trig, radio}$ (\textit{upper}) and particle $r_{\rm trig, part}$ (\textit{lower}) detection, as a function of primary energy and arrival zenith angle for photon (\textit{left}) and proton (\textit{right}) showers.}
    \label{fig:radio_part_trig}
\end{figure}

We quantify the performances of the defined hybrid setup by calculating its trigger efficiency, i.e., the probability of a shower to trigger the array following the criteria defined in Section~\ref{sec:setup}. Namely, the trigger efficiency at $(E,\theta)$ is estimated as $\tau_{\rm trig}(E,\theta)=N_{\rm trig}(E,\theta)/{N_{\rm tot}(E,\theta)}$, where $N_{\rm trig}(E,\theta)$ is the total number of triggered events at energy $E$ and zenith angle $\theta$, and ${N_{\rm tot}(E,\theta)}$ is the total number of showers simulated at $(E, \theta)$. 
Figure~\ref{fig:radio_part_trig} shows the radio and particle trigger efficiencies, over the targeted primary particle parameter range in energy and arrival zenith angle.

While the radio trigger efficiency is close to $100\%$ over the whole parameter range considered for both proton and photon showers, the particle trigger efficiency exhibits a significant difference between these two primaries. Indeed, the muonic component is crucial for the particle detection of very inclined air showers. For hadronic primaries, like protons, the important muon component produced by the decay of charged pions and kaons increases the trigger efficiency, compared to photon primaries for which the muon component can be up to 10 times lower on average \cite{ROS201310,GONZALEZ202048,AbdulHalim:2023Cj,AbdulHalim_2025}, as shown in Figs~\ref{fig:proton_events} and \ref{fig:gamma_events}.

Figure~\ref{fig:radio_part_trig} indicates how particle triggering is intrinsically challenging. Indeed, as mentioned in Section~\ref{sec:signal} and illustrated in Fig.~\ref{fig:profiles_notscaled}, for very inclined showers, the number of particles reaching the ground is lower, and the surviving particles are spread out over a large area --- the typical extent of a shower scales in $1/\cos\theta$, and reaches $\sim 30\,\textrm{km}$ for a zenith angle of $85^\circ$ (see Section~\ref{sec:signal} and \cite{Aab_2018}) --- making the particle trigger efficiency drop at high zenith angles.

Detecting events in radio and then searching for a coincidental signal in particles could be a more efficient approach for enhanced overall detection and cosmic-ray/gamma-ray discrimination. We define the conditional trigger ratio as the probability for the array to be triggered in particle, knowing it was triggered in radio, i.e the ratio of the hybrid trigger efficiency over the radio trigger efficiency, or: $r_{\rm cond} (E, \theta) = N_{\rm trig, hybrid}(E, \theta) /N_{\rm trig, radio}(E, \theta) $, with $N_{\rm trig, hybrid}(E, \theta)$ the number of events triggered both in radio and particles, and $N_{\rm trig, radio}(E, \theta)$ the total number of events triggered in radio.

\begin{figure}
    \centering
    \includegraphics[width=\linewidth]{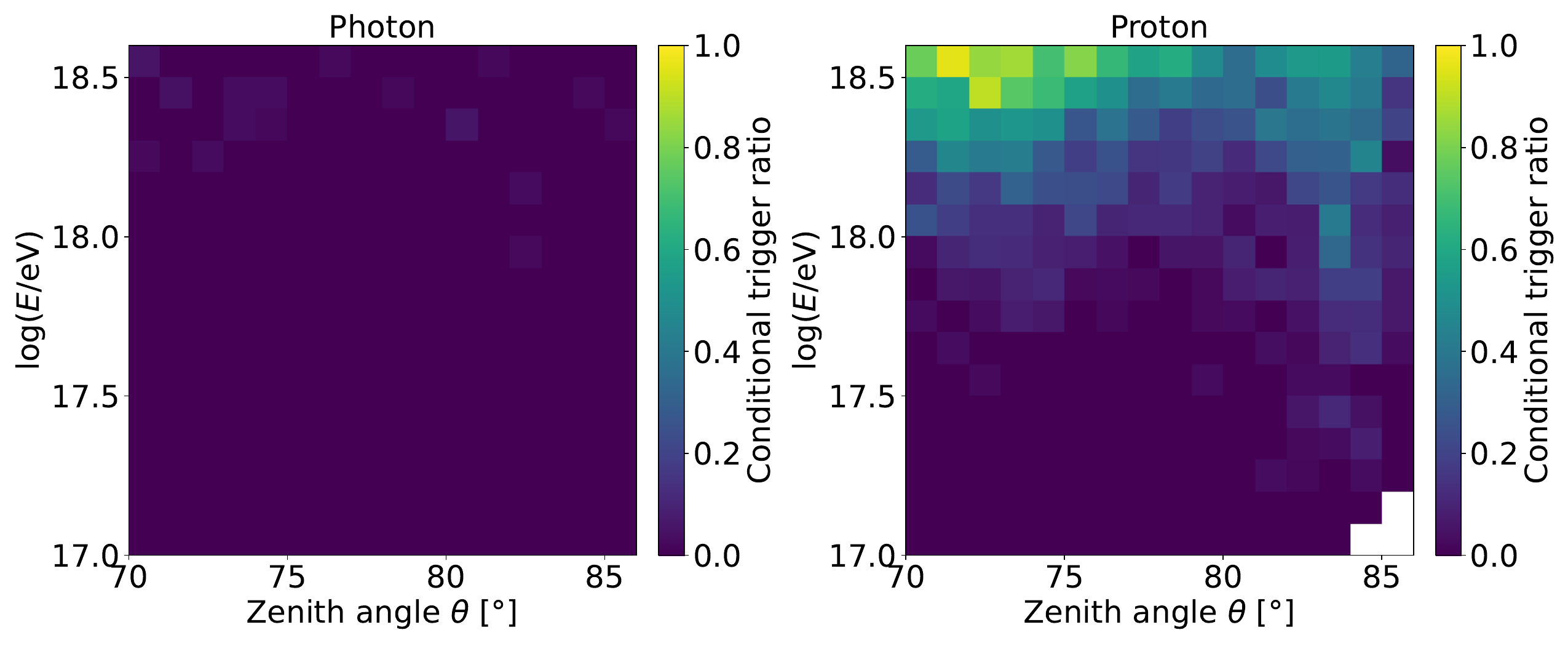}
    \caption{Conditional trigger ratio $r_{\rm cond}$ (ratio of hybrid radio+particle detected events over all radio detected events) as a function of primary energy and arrival zenith angle for photon (\textit{left}) and proton (\textit{right}) showers.}
    \label{fig:cond_trig}
\end{figure}

The conditional trigger ratio for photon and proton showers is presented in Figure~\ref{fig:cond_trig}. We identify a significant parameter range, especially above $10^{18}\,\textrm{eV}$, for which the conditional trigger ratio of proton showers is considerably higher than for photon showers. It underlines the potential of efficiently distinguishing between photon and proton showers with combined radio antennas and scintillation detectors. We quantify this potential in Section~\ref{sec:discr}.

\subsection{Effect of tilting the scintillation detector on detection performances}

\begin{figure}
    \centering
    \includegraphics[width=0.47\linewidth]{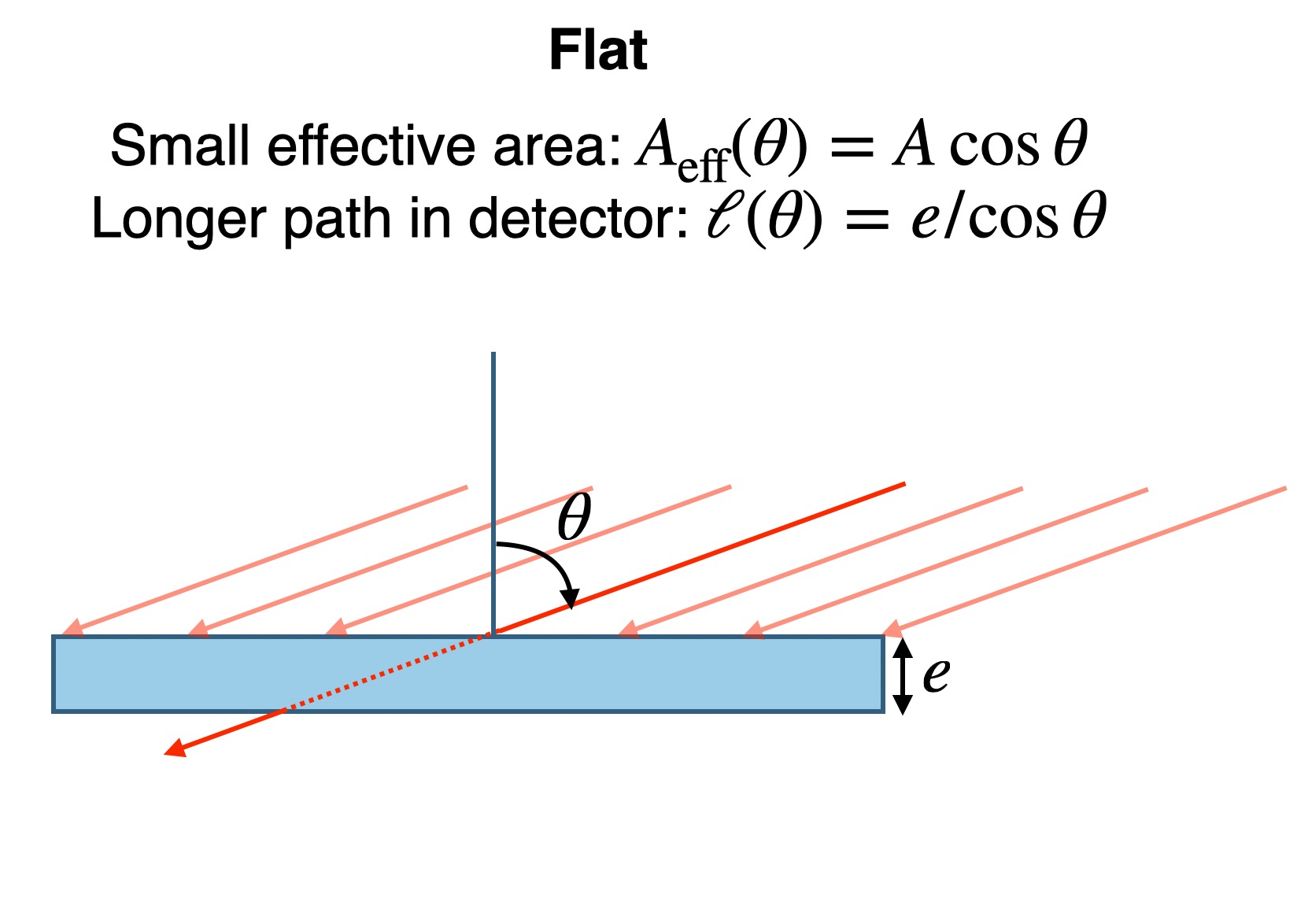}
    \includegraphics[width=0.47\linewidth]{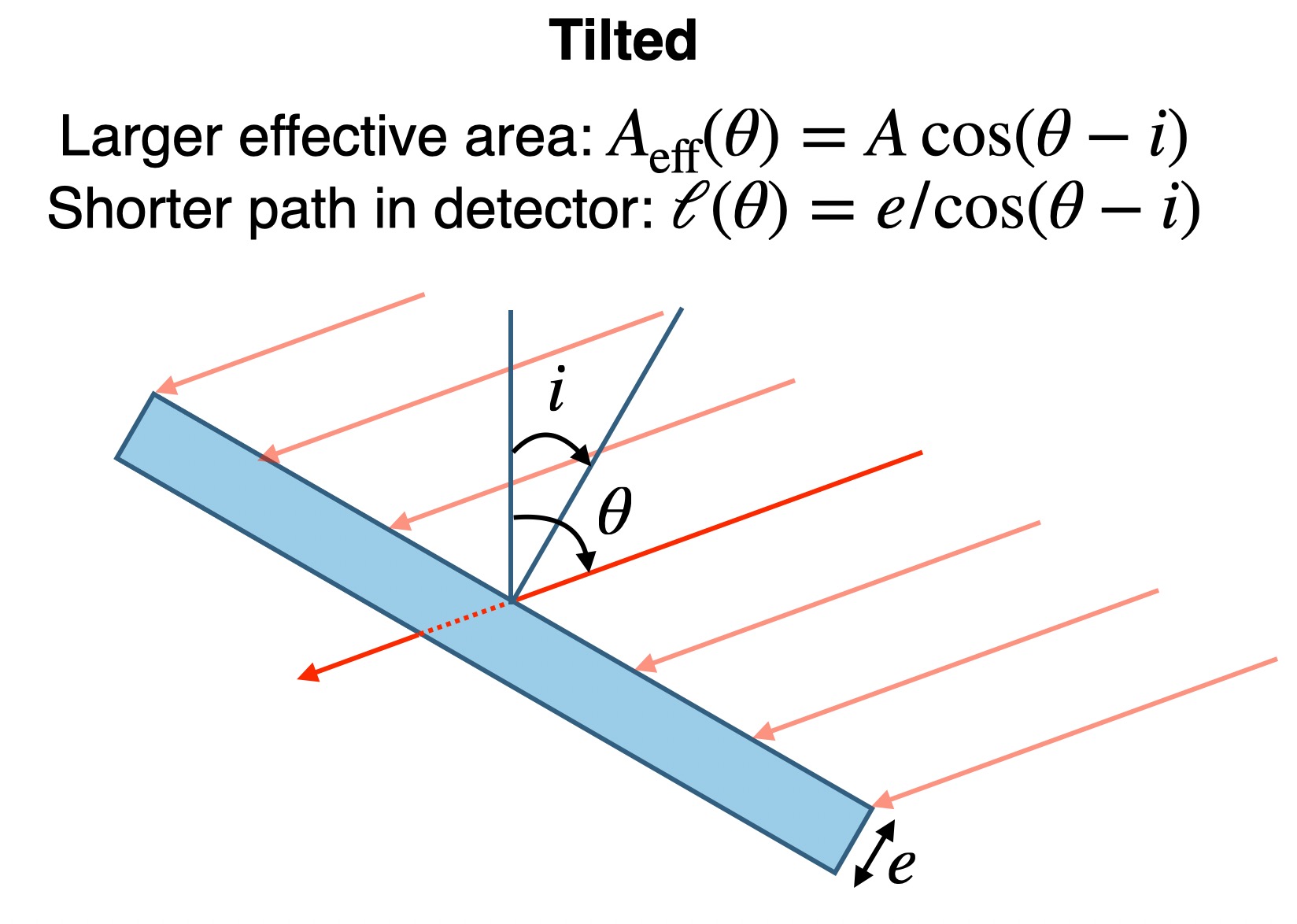}
    \caption{Schematic depiction of a scintillator in the flat (\textit{left}) and tilted (\textit{right}) cases. Both the effective area and the track length of particles vary when inclining the scintillator. These two effects cancel out, leading to an energy deposit in the scintillator independent of the tilt angle $i$.}
    \label{fig:inclined_scint}
\end{figure}

Because the bulk of shower particles reach the ground with zenith angles close to the primary particle inclination, it is commonly inferred for very inclined air-showers that tilting the scintillation detectors would allow for a more efficient detection. 
However, two competing effects are to be considered in the detection process by a scintillator of surface $A$ and thickness $e$ (Figure~\ref{fig:inclined_scint}):
\begin{itemize}
    \item the effective area of the scintillator scales as $A_{\rm eff}(\theta) = A\cos{\theta}$, with $\theta$ the particle zenith angle, corresponding to the angle between the scintillator and the incident particle,
    \item the track length of particles through the scintillator scales approximately as $\ell(\theta) = e/\cos{\theta}$.
\end{itemize}

\begin{figure}
    \centering
    \includegraphics[width=0.5\linewidth]{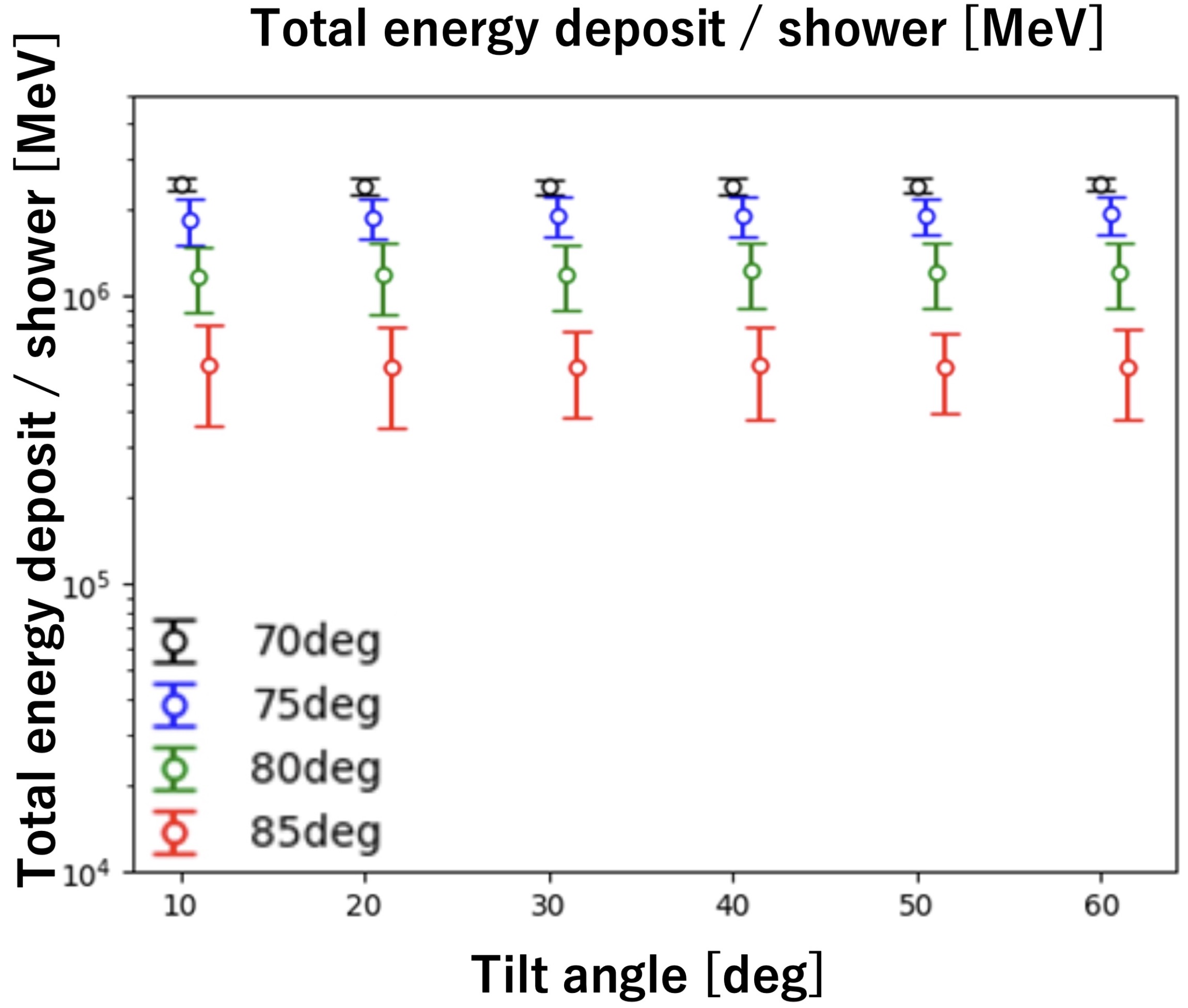}
    \caption{Tilting the scintillation detectors does not affect detector performances for very inclined air-showers. The total energy deposit per shower is shown for simulations performed for 6 scintillator tilt angles, and for various primary arrival zenith angles, as indicated.}
    \label{fig:tilt}
\end{figure}

These scalings yield the average energy deposit of particles in the scintillator:
\begin{equation}
    E_{\rm scint} = \rho A_{\rm eff}(\theta) \, \ell(\theta)\varepsilon = \rho A e \varepsilon \ ,
\end{equation}
with $\varepsilon$ the energy deposit per unit length for minimum ionizing particles (MIPs), and $\rho$ the particle flux on the surface. The total energy deposit does not depend on the scintillator tilt angle. Figure~\ref{fig:tilt} confirms this calculation with air-shower simulations performed for 6 scintillator tilt angles between $[10^\circ,60^\circ]$ and for various primary arrival zenith angles, as indicated. Tilting the scintillation detectors is therefore not necessary to improve the performances of particle detection. However, shower-to-shower fluctuations can lead to large variations of the signal.

\section{Photon and proton-induced shower discrimination method}\label{sec:discr}

Based on previous UHE photon search methods \cite{PhysRevLett.123.051101,2021ExA....52...85K,ABBASI20198,Aab_2017,Abreu_2022,PhysRevD.110.062005,AbdulHalim_2025}, we propose a classifier for the hybrid detection of photons with radio and particle detectors, using two estimators: the total energy deposit of shower particles and the total root mean square of radio traces.

\subsection{Principle}\label{sec:principle}

For each event (proton or photon), we compute two observables, the total energy deposit $E_{\textrm{dep,tot}}$ and the total RMS of the radio traces $\textrm{RMS}_{\rm radio,tot}$, with the following procedure. If the event does not reach the event-level trigger, i.e., less than $N_{\rm ant} = 4$ antennas with a peak-to-peak amplitude above $E_{\rm thr} = 75 \:\mu \textrm{V/m}$, the event is discarded.
For the remaining radio-triggered events, we select all antennas with a peak-to-peak amplitude above the threshold $E_{\rm thr} = 75 \:\mu \textrm{V/m}$, and all scintillators with an energy deposit $\epsilon_{\textrm{dep}}$, averaged between the upper and lower layers, above a threshold of $\epsilon'_{\rm thr} = 0.3\,\textrm{MIP}$. The procedure enables the estimation of the key quantities:
\begin{align}
    \textrm{RMS}_{\rm radio,tot} &= \sum \bigg[\sum_{j=x, y, z}\frac{1}{N}\sum_{i=0}^N E_j^2(t_i)\bigg]^{\frac{1}{2}} = \sum \bigg[\sum_{j=x, y, z}\textrm{RMS}^2_j\bigg]^{\frac{1}{2}}\ ,\\
    E_{\textrm{dep,tot}} &= \sum \epsilon_{\textrm{dep}} = \sum \frac{1}{2}(\epsilon_{\textrm{dep, upper}} + \epsilon_{\textrm{dep, lower}}) \ .
\end{align}

\begin{figure}
    \centering
    \includegraphics[width=0.7\linewidth]{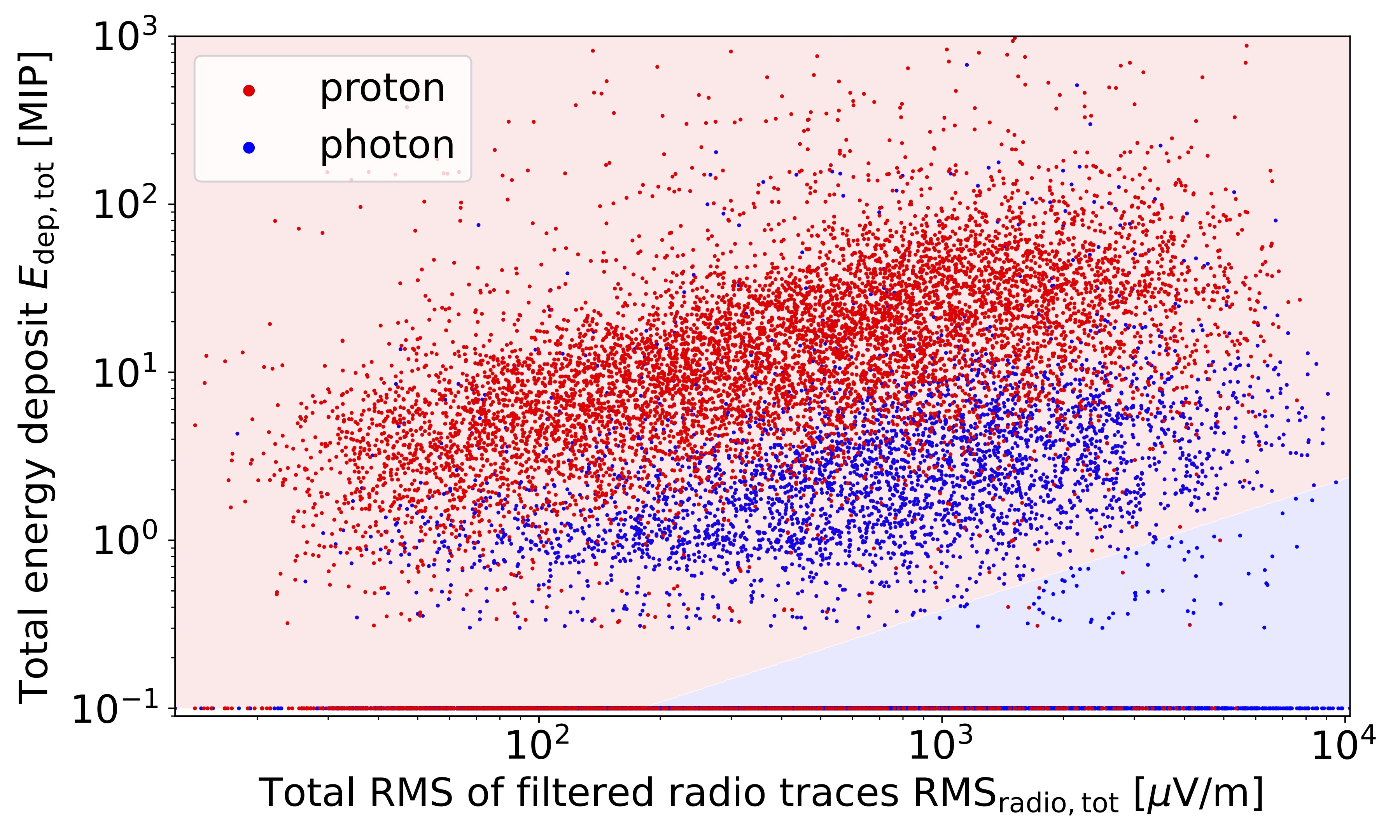}
    \caption{Total energy deposit vs total RMS of the radio traces of triggered antennas, for proton events (red) and photon events (blue). The classification boundary of the logistic regression is also represented. An event in the light red area will be classified as a proton event, and as a photon event in the light blue area. Events with no energy deposit recorded in the scintillators are represented at a total energy deposit of 0.1.}
    \label{fig:RMSvsDE_all}
\end{figure}

Events for which no scintillator reaches the threshold of $\epsilon'_{\rm thr} = 0.3\,\textrm{MIP}$ are assigned the value $E_{\textrm{dep,tot}} = 0.1$. We perform a logistic regression of the selected events, using a linear model in $\log_{10}(E_{\textrm{dep,tot}})-\log_{10}(\textrm{RMS}_{\rm radio,tot})$ space. To account for the low flux of photons when compared to the hadronic background, we weight proton events by a factor of $w_{\rm p} = 10$, to strongly penalize misclassifications of proton events as photon events. Moreover, individual events are weighted with a power-law spectrum $w(E) = E^{- \Gamma}$ of spectral index $\Gamma = 2$ \cite{Abreu_2022, Aab_2017} to account for the expected flux. We denote by $p_\gamma$ the classifier output, interpreted as the predicted probability that an event is a photon event. Figure \ref{fig:RMSvsDE_all} shows the  distribution of $E_{\textrm{dep,tot}}$ and $\textrm{RMS}_{\rm radio,tot}$ for all events, as well as the classification boundary, set by default at $p_\gamma = 0.5$. Events above the boundary, for which $p_\gamma < 0.5$, are classified as proton events, while events below the boundary, with $p_\gamma > 0.5$ are classified as photon events.

We note that the classifier performs poorly on the full batch of simulated events considered without distinction, because of particle-poor proton events. A fraction of proton showers can exhibit a dominant electromagnetic component, generated by the decay of a $\pi^0$ produced in the first hadronic interactions with a significant fraction of the proton energy, resulting in muon-poor events. However, most of these particle-poor events are  events with higher zenith angles, above $80^\circ$, or lower energies, below $10^{17.4} \textrm{ eV}$, for which the particle footprint is either too sparse or too small to be detected (see Figs.~\ref{fig:scatter_energy} and~\ref{fig:scatter_zenith} in Appendix~\ref{sec:appendixA}). 

\subsection{Classification optimization}\label{sec:discr_better}

\begin{figure}
    \centering
    \includegraphics[width=\linewidth]{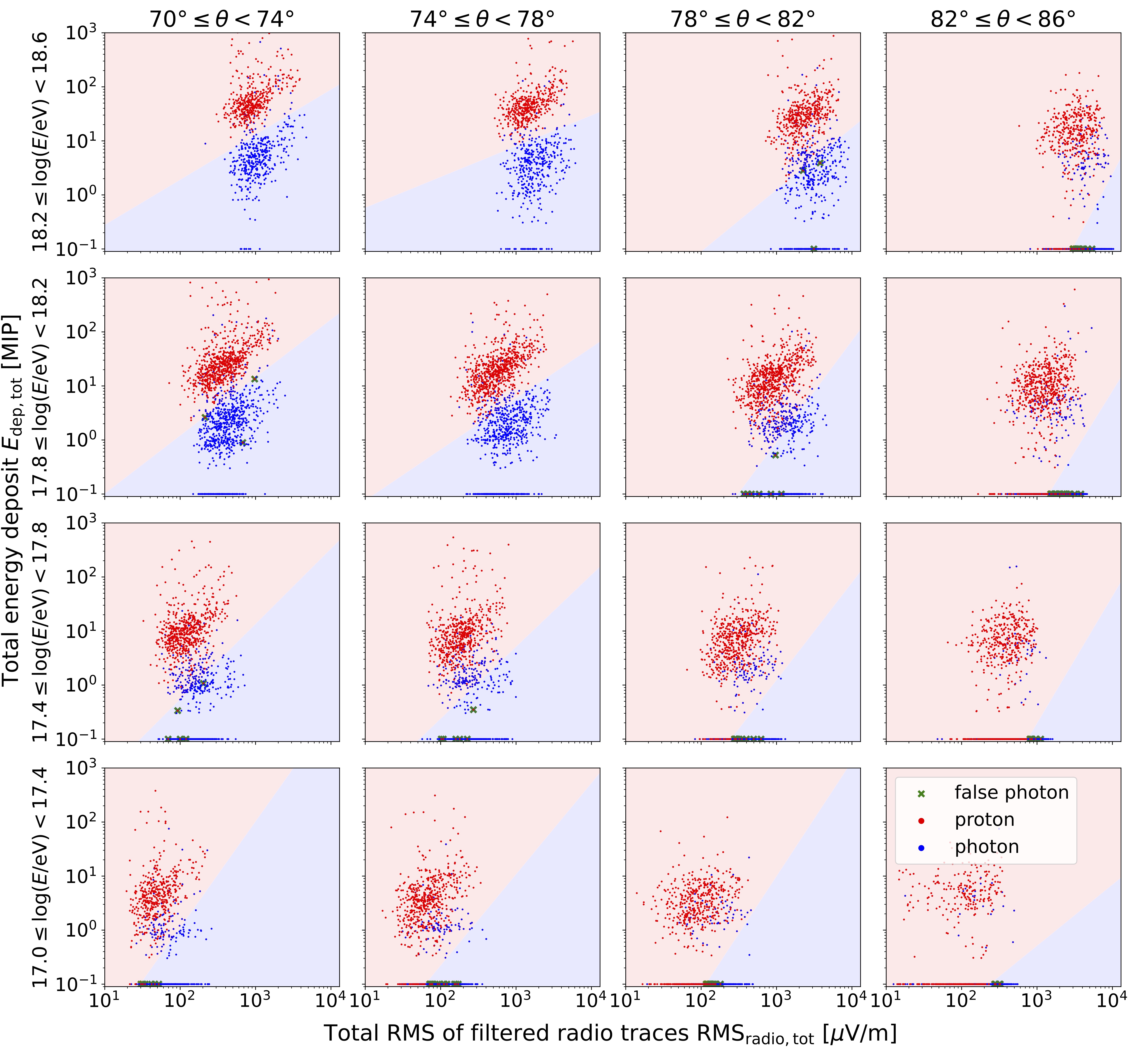}
    \caption{Total energy deposit vs total RMS of the radio traces of triggered antennas, for proton events (red) and photon events (blue), in 16 energy and zenith angle bins. For each bin, an event in the light red area will be classified as a proton event, and as a photon event in the light blue area. Events with no energy deposit recorded in the scintillators are represented at a total energy deposit of 0.1. Proton events falsely identified as photons are shown as green marks.}
    \label{fig:RMSvsDE_binned}
\end{figure}

This energy and zenith dependency is taken into account by splitting our set of simulated events in 4 energy bins of width $\Delta\log_{10}(E/\rm eV) = 0.4$ and 4 zenith angle bins of width $\Delta\theta=4^\circ$, for a total of 16 bins. We repeat the previous analysis in each bin. Figure \ref{fig:RMSvsDE_binned} shows the  distribution of $E_{\textrm{dep,tot}}$ and $\textrm{RMS}_{\rm radio,tot}$ in each bin, as well as the classification boundary at $p_\gamma = 0.5$. We notice a significant improvement compared to the previous case: the number of misclassified proton events is much lower, especially for the higher energy and lower zenith angle bins.

\begin{figure}
    \centering
    \includegraphics[width=0.7\linewidth]{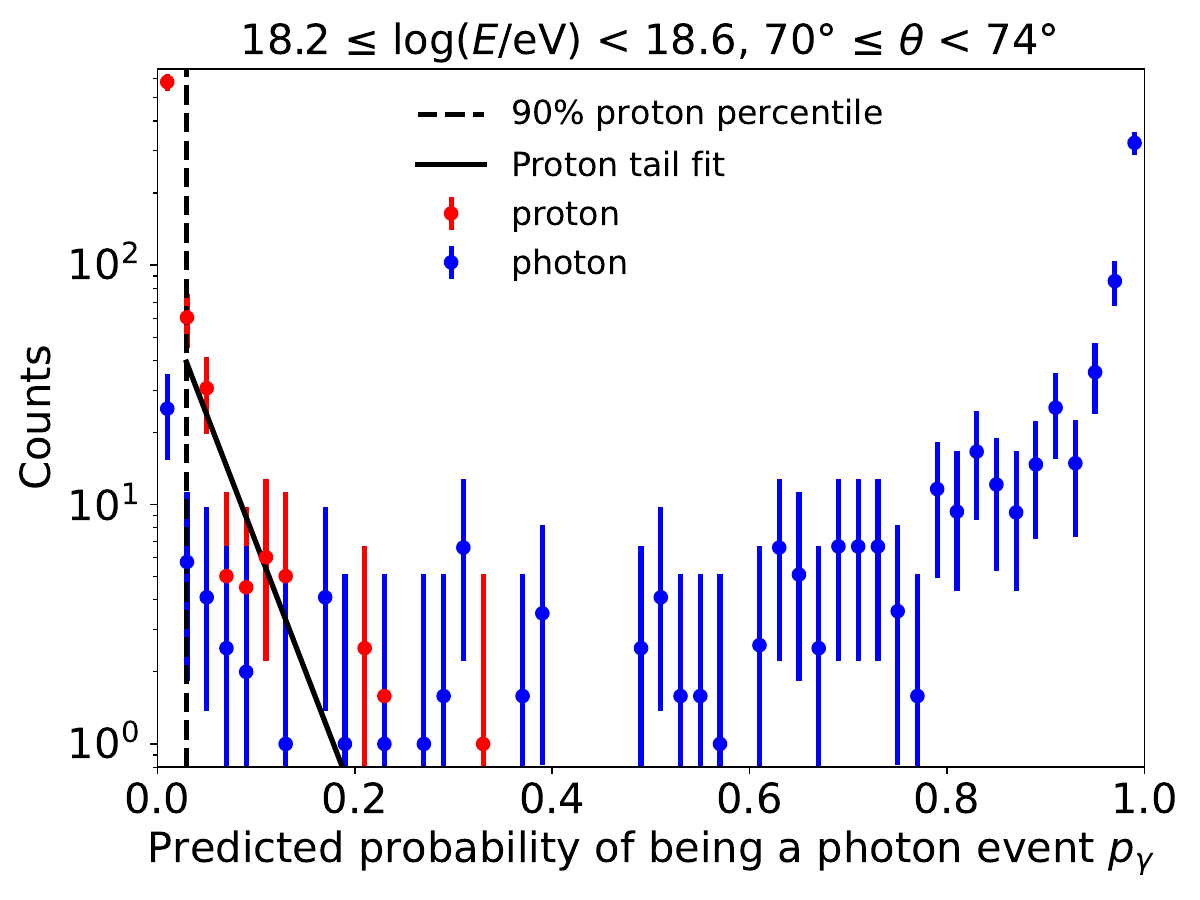}
    \caption{Distribution of the predicted probability of being a photon event $p_\gamma$ returned by the classifier, for the simulated events with energies $18.2 \leq \log_{10}{(E/\textrm{eV})} < 18.6$ and zenith angles $74^{\circ} \leq \theta < 78^{\circ}$. Error bars represent the uncertainty: Poisson errors are used in bins with more than 10 counts, and asymmetric Feldman-Cousins intervals \protect{\cite{PhysRevD.57.3873}} at $95\%$ confidence level are used otherwise (same as \protect{\cite{AbdulHalim_2025}}). The fit of the proton tail is shown as a black line.}
    \label{fig:tail_fit}
\end{figure}

In order to study the classification performance, especially to reduce the background contamination, we vary the cut $p_{\gamma,\textrm{cut}}$ on the predicted probability for an event to be a photon event $p_\gamma$, between $0$ and $1$. An event is classified as a photon event if $p_\gamma > p_{\gamma,\textrm{cut}}$. Signal efficiency is defined as the ratio of the number of photon events correctly classified as photon events, over the total number of photon events. Background contamination is defined as the ratio of the number of proton events incorrectly classified as photon events, over the total number of proton events. Because of the limited statistics in each bin, the contamination calculated this way is discrete. For a more realistic estimation, we perform an unbinned maximum likelihood fit of the tail of the proton $p_\gamma$-distribution. We fit the $10\%$ proton events with higher $p_\gamma$, assuming an exponential tail, following the same procedure as in \cite{AbdulHalim_2025}. In Figure~\ref{fig:tail_fit}, the $p_\gamma$-distributions for proton and photon events with $18.2 \leq \log_{10}(E/\textrm{eV}) < 18.6$ and $74^{\circ} \leq \theta < 78^\circ$ are represented, with the fitted tail of the proton distribution as a black line. The number of misclassified protons is then obtained by integrating the fitted tail between $p_{\gamma,\textrm{cut}}$ and $1$.

\begin{figure}
    \centering
    \includegraphics[width=\linewidth]{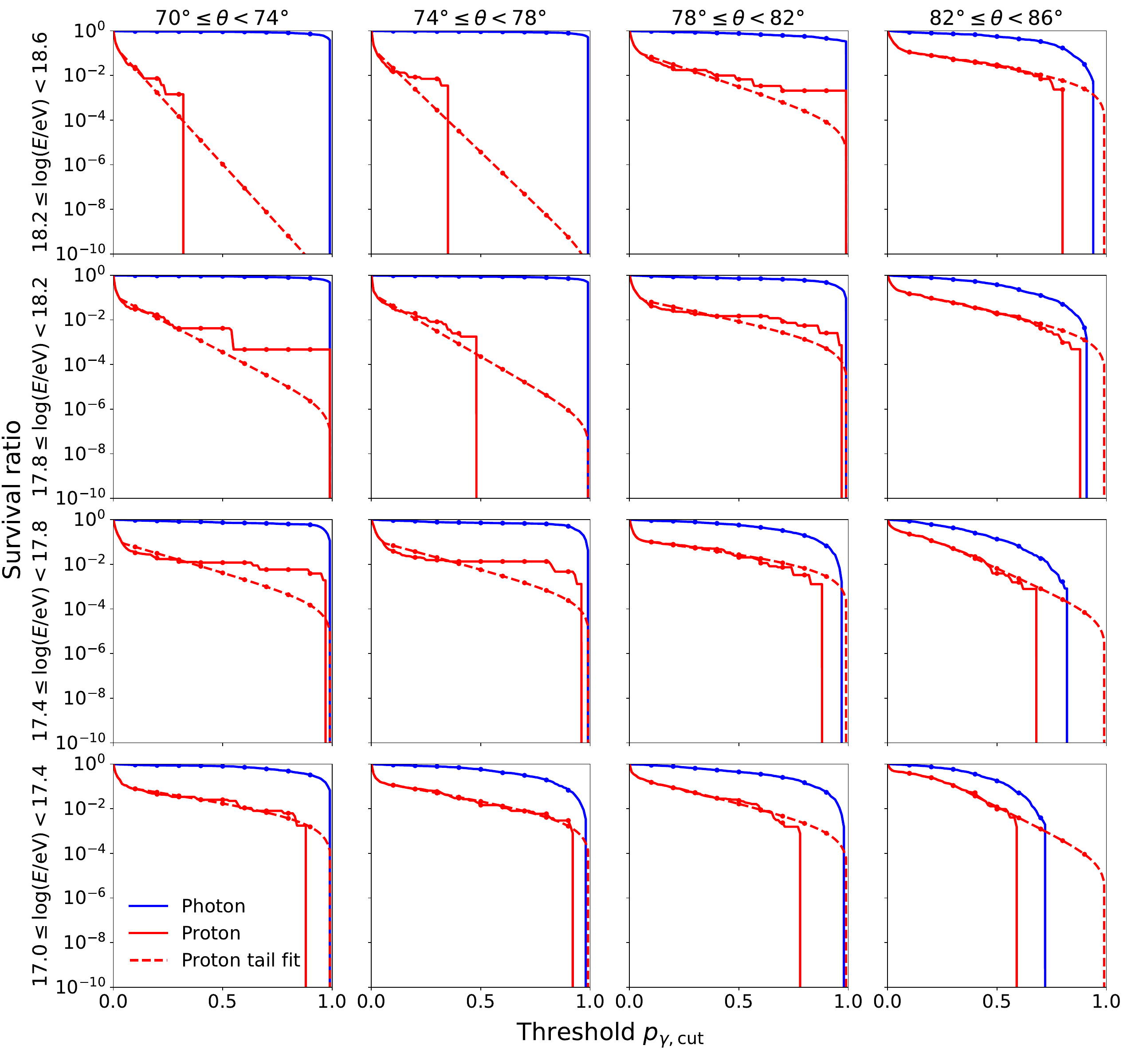}
    \caption{Survival ratio for protons (red) and photons (blue) as a function of the discrimination threshold, in the different energy and zenith angle bins. Solid and dashed lines correspond respectively to the discrete and interpolated ratios. Markers are placed at $p_{\gamma,\rm cut}$ threshold values multiple of $0.1$.}
    \label{fig:survival_vs_thresh}
\end{figure}

Figure \ref{fig:survival_vs_thresh} shows the signal efficiency and background contamination, in both the discrete and interpolated cases, computed by scanning over all possible values of $p_{\gamma,\textrm{cut}}$, for all energy and zenith angle bins. Markers are placed at threshold values multiple of $0.1$. Increasing the $p_{\gamma,\textrm{cut}}$ threshold allows to reduce the background contamination, at the cost of a lower signal efficiency. For the high zenith angle bins, above $82^\circ$, the signal efficiency for photon events drops to $0$, before the background contamination, indicating that the particle footprint of the shower becomes too sparse for a sufficient sampling.

\begin{figure}
    \centering
    \includegraphics[width=\linewidth]{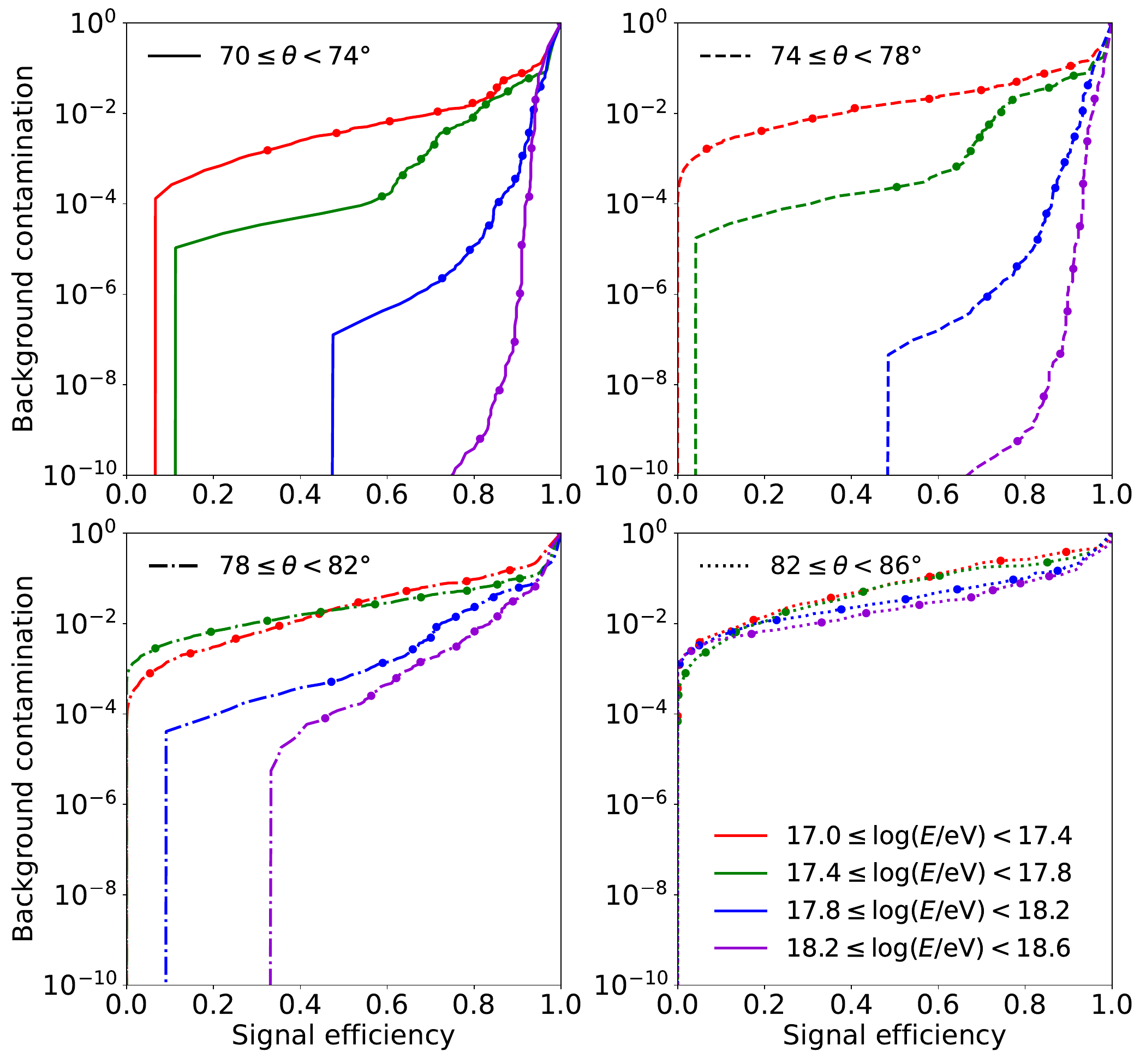}
    \caption{Background contamination in terms of signal efficiency for various zenith angle bins. Energy bins are color-coded. Markers are placed at $p_{\gamma,\rm cut}$ threshold values multiple of 0.1.}
    \label{fig:thresh_vars}
\end{figure}

The interpolated background contamination is represented as a function of the signal efficiency on Figure~\ref{fig:thresh_vars}. In general, at a given signal efficiency, the background contamination is lower at higher energies, because of the higher number of muons in proton showers at higher energies, and at lower zenith angles, because of the improved sampling of the particle footprint of the showers. For the default classification threshold $p_{\gamma,\rm cut} = 0.5$, the background contamination varies significantly depending on the energy or zenith angle bin, between a minimum of $10^{-4} \%$ for high energies and low zenith angles, and a maximum of $2.66\%$ at lower energies and higher zenith angles, while the signal efficiency varies between $13.4\%$ and $91.1\%$. Choosing a threshold $p_{\gamma,\textrm{cut}}=0.9$ allows to reach a background contamination lower than $2.5\times10^{-3}$ in all bins, at the cost of a lower signal efficiency, between $0\%$ and $78.2\%$. The choice of an optimal $p_{\gamma,\rm cut}$ is described in Section~\ref{sec:flux} and in more details in Appendix~\ref{sec:appendixB}.

\subsubsection{Impact of reconstruction resolution}\label{sec:z_only}

\begin{figure}
    \centering
    \includegraphics[width=\linewidth]{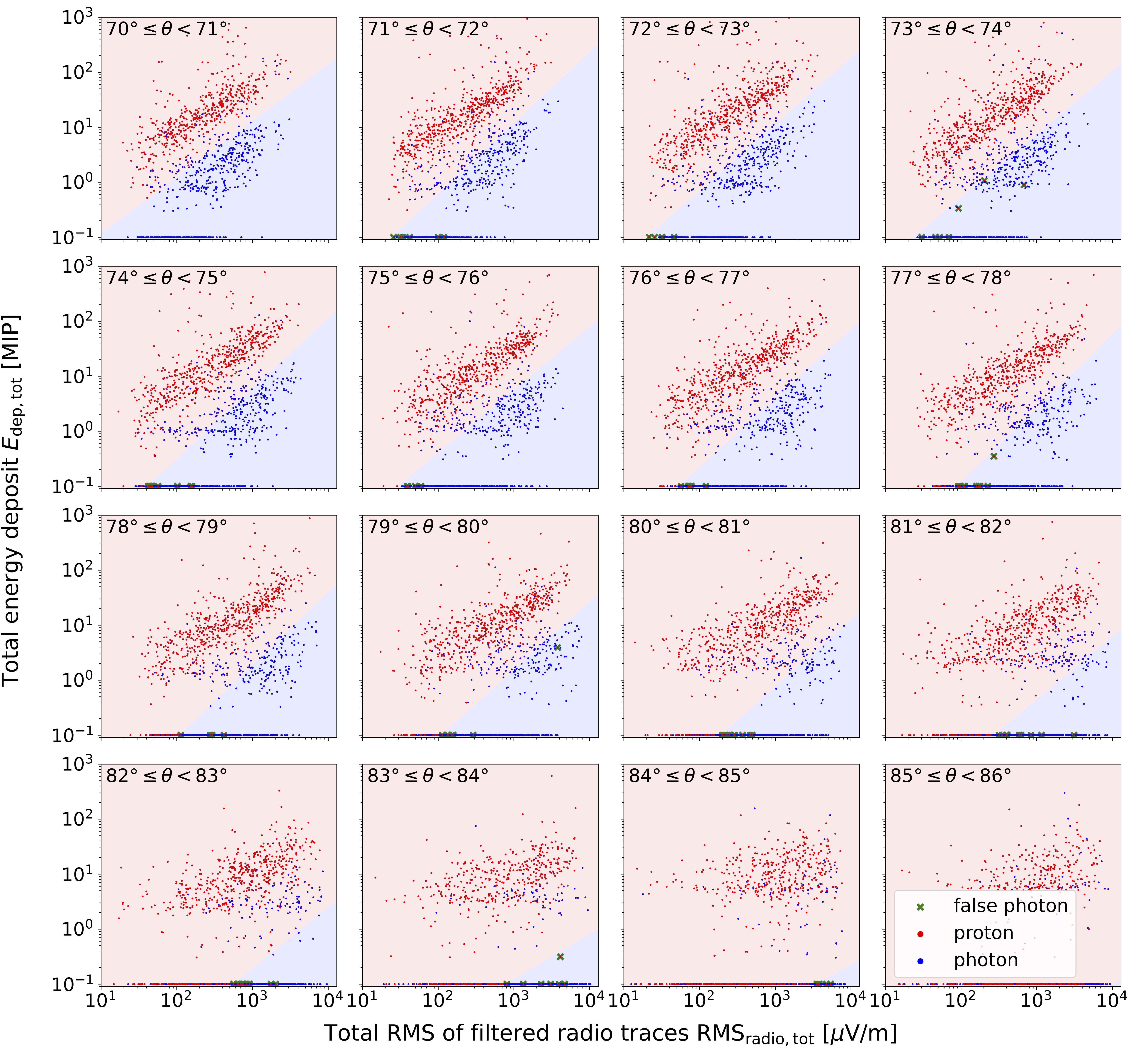}
    \caption{Same as Figure~\ref{fig:RMSvsDE_binned}, in 16 zenith angle bins of width $\Delta\theta=1^{\circ}$, with uniform sample weights $w(E)=1$.}
    \label{fig:RMSvsDE_z_only}
\end{figure}

To perform this analysis, we must assign an energy $E$ and zenith angle $\theta$ to each event. The reconstruction of the energy of the primary and its arrival direction from the radio signal measured by the antennas is currently being investigated for GRAND, using various approaches \cite{Zhang:2025n3,Guelfand:2025lY,Gulzow:2025iV,Ferriere:2025FK,Benoit-Levy:2025/j}. In the absence of a definitive procedure for now, we use the true energy and zenith angle of the shower, as extracted from the simulations. Although the radio signal can be used to reconstruct the electromagnetic energy of the shower with an intrinsic resolution of $\sim 5\%$ \cite{GUELFAND2025103120}, fluctuations due to either the primary particle type or the initial stage of the development of the shower can lead to significant uncertainties on the reconstructed primary energy. However, the total RMS of the filtered radio traces $\textrm{RMS}_{\rm radio,tot}$ is strongly correlated to the primary energy, allowing to keep sufficient information on the energy. On the other hand, the primary zenith angle can be reconstructed with a resolution of $\sim 0.1^\circ$ with current methods \cite{GUELFAND2025103120}.
To evaluate the influence of energy reconstruction resolution, we tested configurations without energy binning. Additionally, to assess the effect of zenith reconstruction resolution, we considered two zenith binning configurations: broad bins with $\Delta\theta = 4^\circ$ and thin bins with $\Delta\theta = 1^\circ$, both covering the range $70^\circ \leq \theta < 86^\circ$. As described in Sec.~\ref{sec:discr}, events are weighted with $w(E) \propto E^{-2}$ to account for the expected photon flux. However, removing energy binning increases the energy range per bin, leading to higher contamination at high energies (high $\mathrm{RMS}_{\mathrm{radio,tot}}$) and degraded average classification performance. Omitting the sample weighting when fitting the boundary improves the classification, particularly for high energies events. Therefore, for each zenith binning configuration, we test both flux-proportional weights, $w(E) \propto E^{-2}$, and uniform weights, $w(E) = 1$.
Figure~\ref{fig:RMSvsDE_z_only} shows the classification boundary at $p_\gamma = 0.5$ in the $E_{\mathrm{dep,tot}}$ vs. $\mathrm{RMS}_{\mathrm{radio,tot}}$ space for $\Delta\theta = 1^\circ$, uniform weights $w(E) = 1$, and no energy binning. While this configuration allows for a better classification at high $\mathrm{RMS}_{\mathrm{radio,tot}}$, it does so at the expense of signal efficiency. For the flux limit calculations, we retain the original energy and zenith binning. The consequences of removing energy binning on the flux limits are discussed in detail in Appendix~\ref{sec:appendixC}.

\subsection{Exposure and flux limits}\label{sec:flux}

\begin{figure}
    \centering
    \includegraphics[width=\linewidth]{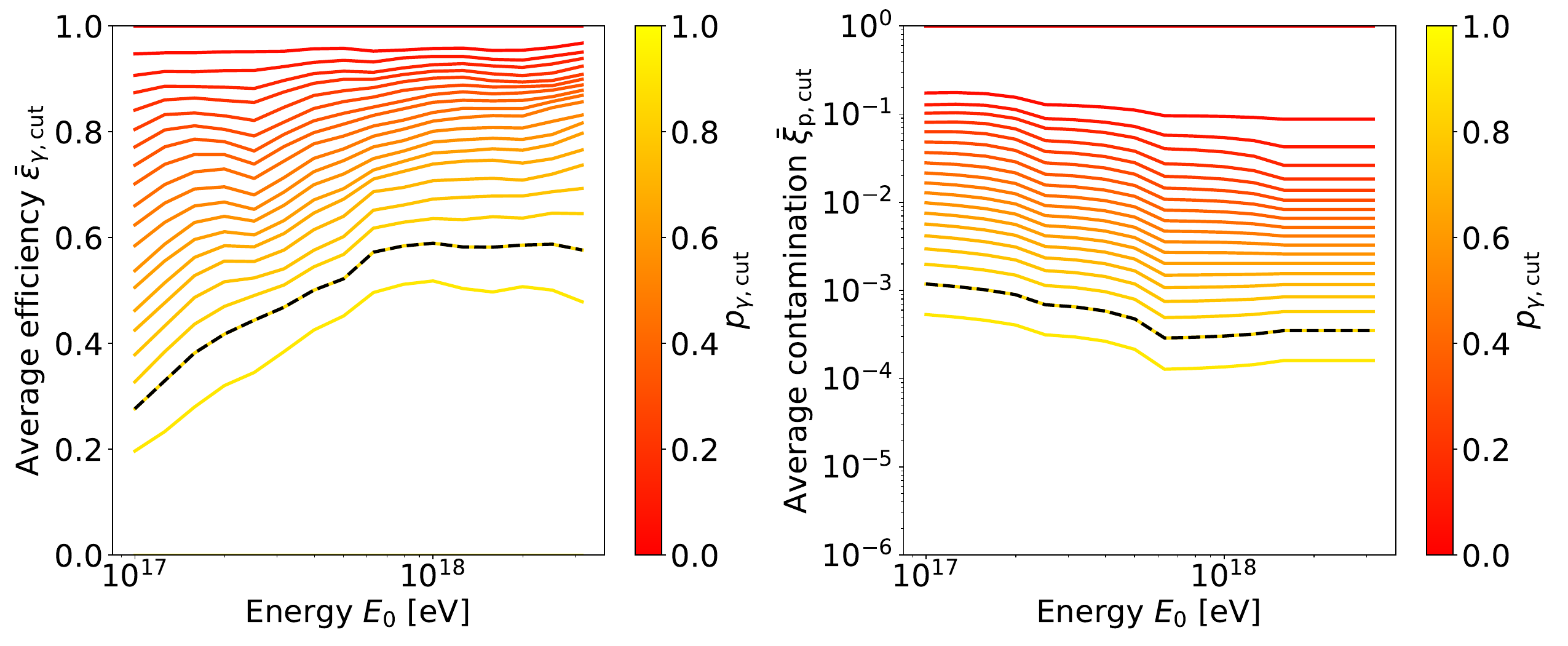}
    \caption{Average signal efficiency $\bar{\varepsilon}_{\gamma,\rm cut}$ and background contamination $\bar{\xi}_{\rm p, cut}$ as a function of the energy $E_0$, with the threshold $p_{\gamma, \rm cut}$ color-coded. The optimal value $p_{\gamma, \rm cut}=0.9$ is identified with a dashed black line.}
    \label{fig:effi_conta}
\end{figure}

In order to determine the optimal value of $p_{\gamma,\rm cut}$, we compute the average signal efficiency $\bar{\varepsilon}_{\gamma,\rm cut}$ and background contamination $\bar{\xi}_{\rm p, cut}$ of the photon candidate cut described in Section~\ref{sec:discr_better} above an energy $E_0$. It is calculated by integrating the signal efficiency and background contamination with weights proportional to the energy spectrum and the arrival direction:
\begin{equation}\label{eq:effi}
    \bar{\varepsilon}_{\gamma,\rm cut}(E_\gamma>E_0) 
    = \frac{\int_{E_0}^{\infty}\int_0^{\frac{\pi}{2}}E_\gamma^{-\Gamma}\tau(E_\gamma, \theta) \varepsilon_{\gamma,\textrm{cut}}(E_\gamma,\theta)\cos{\theta} \sin{\theta} \,\textrm{d}\theta \, \textrm{d}E_\gamma}{\int_{E_0}^{\infty}\int_0^{\frac{\pi}{2}}E_\gamma^{-\Gamma}\tau(E_\gamma, \theta) \cos{\theta} \sin{\theta} \,\textrm{d}\theta \, \textrm{d}E_\gamma}\ ,
\end{equation}
and similarly, for the average background contamination:
\begin{equation}\label{eq:xi}
    \bar{\xi}_{\rm p, cut}(E_{\rm p}>E_0) = \frac{\int_{E_0}^{\infty}\int_0^{\frac{\pi}{2}}E_{\rm p}^{-\Gamma_{\rm p}}\tau_{\rm p}(E_{\rm p}, \theta) \xi_{\textrm{p,cut}}(E_{\rm p},\theta)\cos{\theta} \sin{\theta} \,\textrm{d}\theta \, \textrm{d}E_{\rm p}}{\int_{E_0}^{\infty}\int_0^{\frac{\pi}{2}}E_{\rm p}^{-\Gamma_{\rm p}}\tau_{\rm p}(E_{\rm p}, \theta) \cos{\theta} \sin{\theta} \,\textrm{d}\theta \, \textrm{d}E_{\rm p}}\ ,
\end{equation}
where $\varepsilon_{\gamma,\textrm{cut}}(E_\gamma,\theta)$ is the efficiency of the photon candidate cut for radio-triggered events, and $\xi_{\textrm{p,cut}}(E_{\rm p},\theta)$ the background contamination, or background survival efficiency.
$\tau(E_\gamma, \theta)$ and $\tau_{\rm p}(E_{\rm p}, \theta)$ are the radio-trigger efficiencies for photon and proton events respectively. The spectral indices used are $\Gamma = 2$ \cite{Abreu_2022, Aab_2017} for photon events and $\Gamma_{\rm p} = 3.3$ for proton events \cite{Abreu_2021}. The average signal efficiency $\bar{\varepsilon}_{\gamma,\rm cut}$ and background contamination $\bar{\xi}_{\rm p, cut}$ calculated as a function of $E_0$ are shown in figure \ref{fig:effi_conta}, for varying values of $p_{\gamma, \rm cut}$. Because the photon flux is expected to be much lower than the cosmic ray flux for the energy range considered, we want to minimize the background contamination, while maintaining a satisfying signal efficiency. We determine that a threshold value $p_{\gamma, \rm cut} = 0.9$ is the best compromise in our case, with a maximum average background contamination of $0.12\%$ (see Appendix~\ref{sec:appendixB} for more details).

The integrated efficiency-weighted exposure for photons with energies above $E_0$, $\mathcal{E}_\gamma(E_\gamma>E_0)$, is computed from simulations as:
\begin{equation}
    \mathcal{E}_\gamma(E_\gamma>E_0) = 2\pi A \Delta T\int_{E_0}^{\infty}\int_0^{\frac{\pi}{2}}\frac{E_\gamma^{-\Gamma}}{c_E}\tau(E_\gamma, \theta) \cos{\theta} \sin{\theta} \,\textrm{d}\theta \, \textrm{d}E_\gamma\ ,
\end{equation}
with $\tau(E_\gamma,\theta)$ the radio-trigger efficiency of photon showers, $A$ the area of the detector layout and $\Delta T$ the operation duration of the detector. $c_E$ is a normalization factor because of the weighting of events following a power-law spectrum proportional to $E^{-\Gamma}$, with $\Gamma=2$ for photons, so as 
\begin{equation}\label{eq:ce}
c_E(\Gamma) = \int_{E_0}^{\infty}E_\gamma^{-\Gamma}\textrm{d}E_\gamma \ .
\end{equation}

Assuming no significant excess is observed, the upper limits on the integral photon flux $\Phi_{\gamma,\textrm{U.L.}}^{\textrm{C.L.}}(E_\gamma > E_0)$ achievable by such a hybrid array of radio antennas and scintillation detectors, at a confidence level C.L. and above an energy $E_0$, then reads
\begin{equation}
    \Phi_{\gamma,\textrm{U.L.}}^{\textrm{C.L.}}(E_\gamma>E_0) = \frac{N_\gamma^{\textrm{C.L.}}}{\bar{\varepsilon}_{\gamma,\textrm{cut}}\mathcal{E}_\gamma} \ ,
    \label{flux_ul}
\end{equation}
with $N_\gamma^{\textrm{C.L.}}$ the upper limit on the number of photon events at confidence level C.L., computed using the Feldman-Cousins method \cite{PhysRevD.57.3873}.

\begin{figure}
    \centering
    \includegraphics[width=0.8\linewidth]{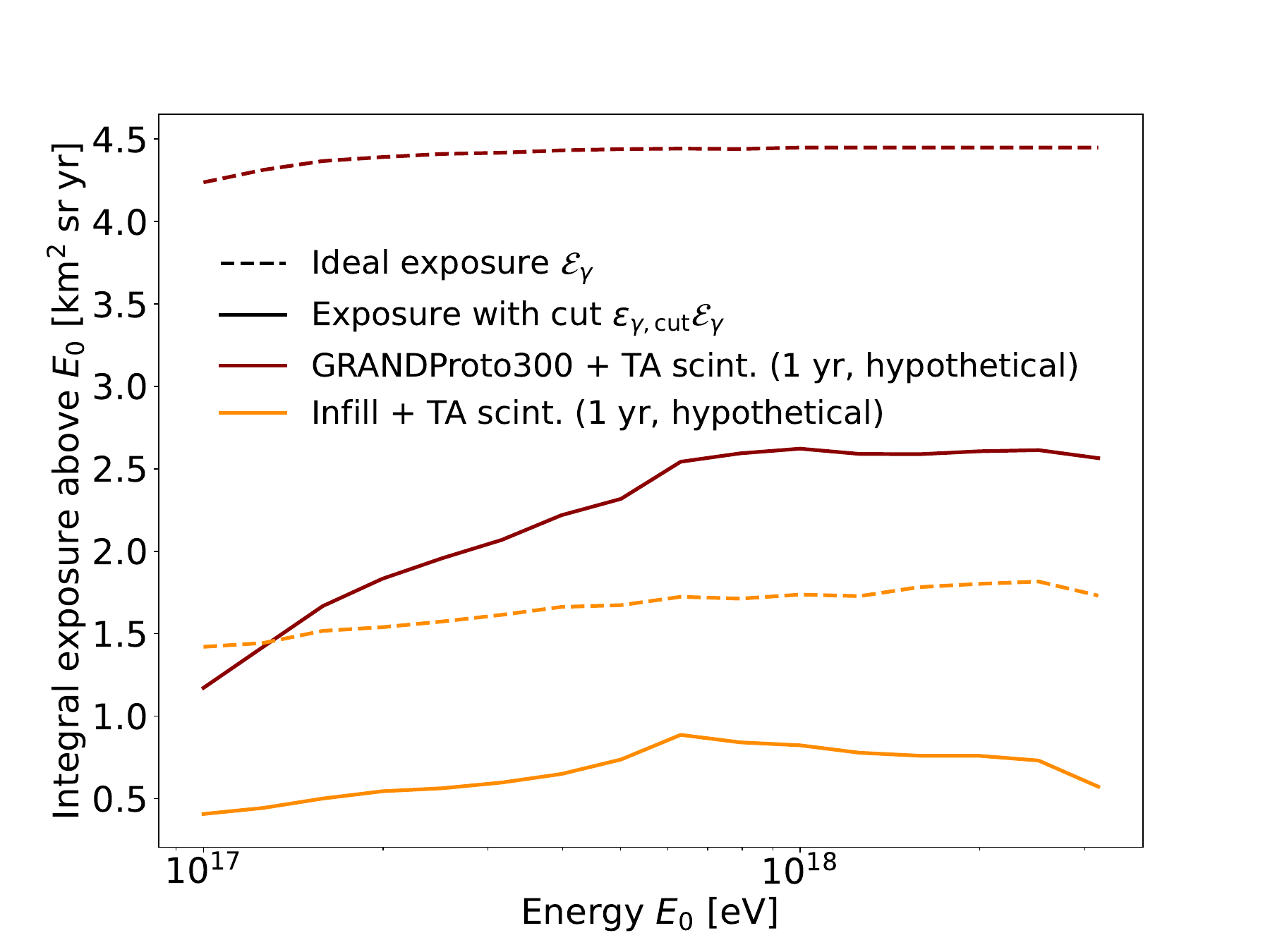}
    \caption{Integrated efficiency-weighted exposure with photon candidate cut $\bar{\varepsilon}_{\gamma,\textrm{cut}}\mathcal{E}_\gamma$ (solid) and ideal exposure $\mathcal{E}_{\gamma}$ (dashed) as a function of the energy for a hypothetical hybrid array of radio antennas and scintillators on the full GRANDProto300 layout and on the infill layout, for an operation duration $\Delta T=1$ yr.}
    \label{fig:exposure}
\end{figure}

\begin{figure}
    \centering
    \includegraphics[width=0.8\linewidth]{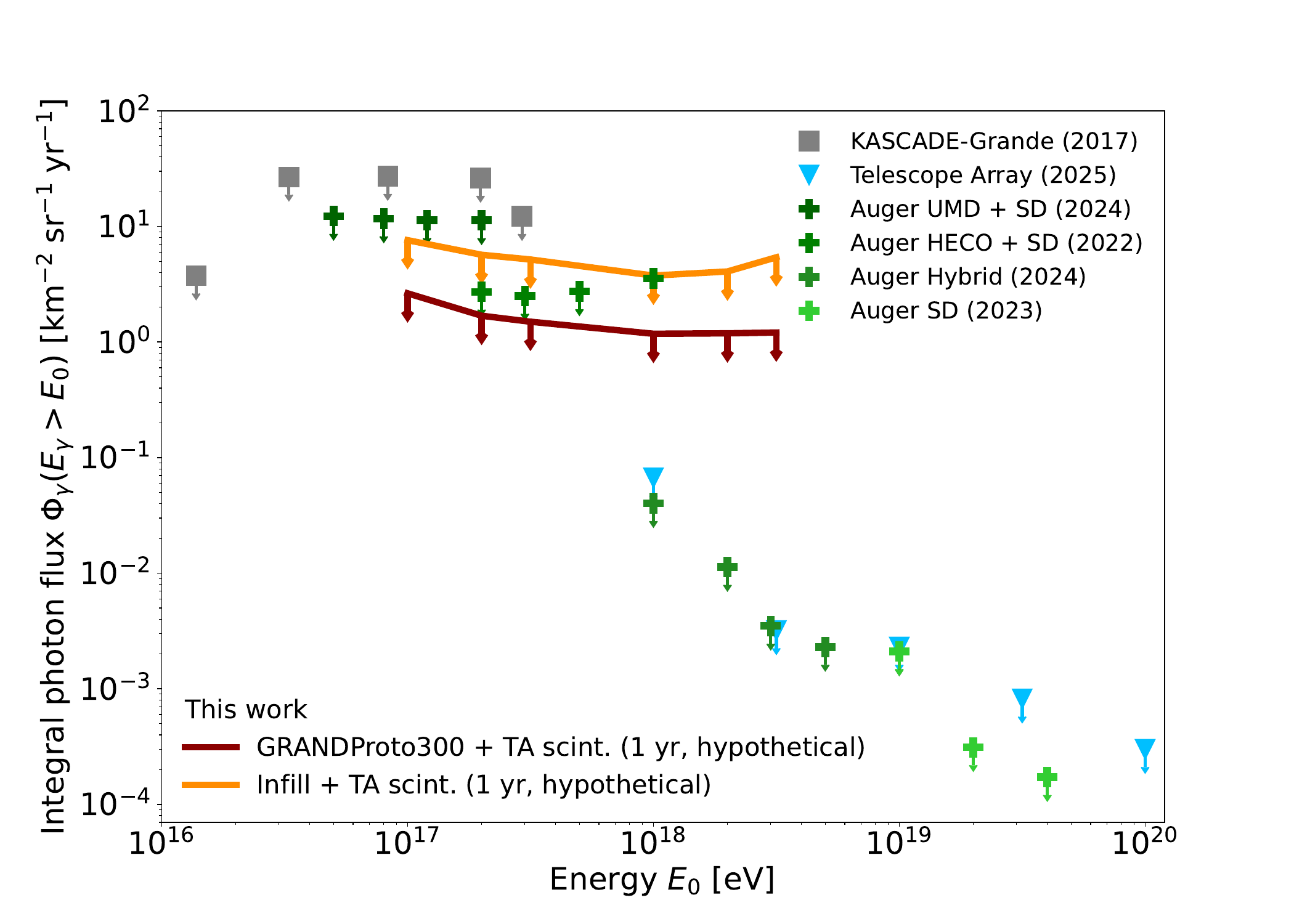}
    \caption{Upper limits (at 95\% C.L.) on the integral photon flux above $10^{17} \textrm{ eV}$ determined in this work, for a hypothetical hybrid array of radio antennas and scintillators on the full GRANDProto300 layout and on the infill layout, for an operation duration $\Delta T=1$ yr. Existing limits above $10^{16}\,\textrm{eV}$ from KASCADE-Grande \protect{\cite{Apel_2017}}, the Pierre Auger Observatory \protect{\cite{Abreu_2023,PhysRevD.110.062005,Abreu_2022,AbdulHalim_2025}} and Telescope Array \protect{\cite{ABBASI20198, Abbasi:2021Z9, PhysRevD.88.112005, 2025arXiv251201638T}} are also reported.}
    \label{fig:flux_ul}
\end{figure}

Both the ideal exposure $\mathcal{E}_{\gamma}$ and the exposure with the photon candidate cut $\bar{\varepsilon}_{\gamma,\textrm{cut}}\mathcal{E}_\gamma$ are shown in Figure~\ref{fig:exposure}, for a hypothetical hybrid array of radio antennas and scintillators on the full GRANDProto300 layout, and on the denser infill array only, assuming an operation duration $\Delta T = 1$ yr. We compute the associated upper limits at $95\%$ confidence level, for no observed photon event, under the conservative hypothesis of no background, which corresponds to $N_\gamma^{95\%} = 3.095$ \cite{PhysRevD.57.3873}. The results are shown in Figure~\ref{fig:flux_ul}, as well as existing limits by KASCADE-Grande \cite{Apel_2017}, the Pierre Auger Observatory \cite{Abreu_2023,PhysRevD.110.062005,Abreu_2022,AbdulHalim_2025} and Telescope Array \cite{ABBASI20198, Abbasi:2021Z9, PhysRevD.88.112005, 2025arXiv251201638T}. Between $10^{17} \,\textrm{eV}$ and $10^{18}\,\textrm{eV}$, the predicted upper limits for a hybrid array of the size of the infill are in the same range as the limits on the integral photon flux obtained by the Pierre Auger Observatory using hybrid data from fluorescence detectors and surface detectors \cite{Abreu_2022}, providing a complementary method to constrain the photon flux in a different zenith range (above $70^\circ$) and using a different detection method, based on radio antennas. Using a larger hybrid array, for example of the size of the full GRANDProto300, would improve the achievable limits by a factor $\sim 3$, with roughly $3$ times more antennas.

\subsection{Discussion and possible improvements}

This study aims at estimating the potential for ultra-high-energy photon identification using a hybrid array of radio antennas and scintillation detectors. Because no such detector currently exists, it is limited by the absence of real data to validate the performances of the method. A properly calibrated reconstruction method for both proton and photon showers from the radio signal is required to provide a reliable estimate of the true energy, zenith and azimuth angles of the primary particle. In the case of GRAND, reconstruction methods of the radio signal are being investigated, using physical modeling \cite{Zhang:2025n3,Guelfand:2025lY,Gulzow:2025iV}, neural networks \cite{Ferriere:2025FK}, and denoising \cite{Benoit-Levy:2025/j}. The reconstruction performances, in particular the relative error on the energy and the zenith angle, will impact the discrimination performances and need to be properly accounted for in future studies. 

Statistics are limited by the number of simulations available (roughly $9000$ events for each primary). Simulating the radio emission of air showers is computationally expensive, because the electric field needs to be calculated at each antenna position (with a linear scaling in the number of antennas). Most air showers experiments re-use simulated showers by randomly displacing the shower core over the detector layout (see \cite{AbdulHalim_2025, PhysRevD.88.112005} for example), effectively multiplying the number of simulated events. However, this method is not applicable for the radio emission, because the electric field is only computed at specified antenna positions. Moreover, in the case of very inclined showers and because of the strength of the magnetic field of the Earth at the site of GRANDProto300 in China, geomagnetic effects become significant and introduce strong variations of the particle energy deposit on ground, depending on the primary particle azimuth angle. 

An additional source of uncertainty stems from the dethinning procedure of Telescope Array \cite{STOKES2012759}, which we used to compute the energy deposit in scintillation detectors. Since TA mainly targets showers with zenith angles lower than $60^\circ$, the procedure was optimized by using a maximal radius of $8.4 \, \textrm{km}$, sufficient to see the whole particle footprint. However, since we target very inclined showers, particle footprint on the grounds can extend over much larger distances. The dethinning code for now does not allow to compute detector response over such distances. Extending the footprint size would be necessary for a more complete analysis.

Finally, the discrimination method presented here uses only two basic observables -- the total RMS of the radio signal and the total energy deposit. Taking into account additional parameters like the number of activated scintillators (see Appendix~\ref{sec:appendixA}), with an energy deposit above the threshold $\epsilon'_{\rm thr} = 0.3 \,\rm MIP$, and performing a multivariate analysis (MVA) could improve the discrimination performances, as demonstrated in similar cases in \cite{Abreu_2022,ABBASI20198}. Defining more robust observables, more sensitive to the muon content of the showers, as in \cite{AbdulHalim_2025}, or applying machine learning methods \cite{Abreu_2022,ABBASI20198,2025arXiv251201638T}, are other axes of improvement. The method could also be applied on other types of hybrid radio and particle detectors arrays, for example with water Cherenkov detectors instead of scintillation detectors, to confirm their potential for UHE photon searches.

\section{Conclusions, perspectives}

The search for UHE photons has so far been conducted with detectors targeting showers with zenith angles lower than $60^{\circ}$, using particle detectors on the ground, complemented with other types of detectors, like fluorescence detectors or underground muon detectors. In this study, we presented a new method to distinguish between photon and proton primaries using a hybrid array of radio antennas and scintillation detectors on the ground. For very inclined air showers, with zenith angles larger than $70^{\circ}$, the hadronic and electromagnetic components of the shower are absorbed before reaching the ground, whereas the muonic component survives. Radio antennas allow us to probe the electromagnetic component of the shower and scintillation detectors its muon content. Because of the lower muon content of photon-induced showers, we distinguish photon and proton events by comparing the total energy deposit of the shower $E_{\rm dep,tot}$ and the total RMS of the radio traces $\textrm{RMS}_{\rm radio,tot}$, using a logistic regression with a linear model in the $\log(E_{\rm dep,tot})-\log(\textrm{RMS}_{\rm radio,tot})$ space, in 16 energy and zenith angle bins. Using the layout of GRANDProto300 as a case study for such a hybrid array, we reach a background contamination below $10^{-4}$ for energies above $10^{17.8}\,\textrm{eV}$ and zenith angles below $82^\circ$. The maximal background contamination is $2.6\%$ for the lowest energies and highest zenith angles. 

We derived the associated upper limits on the integral photon flux between $0.1\,\textrm{EeV}$ and $3\,\textrm{EeV}$, ranging from $~1.2$ to $~2.6\,{\rm km^{-2}\,sr^{-1}\,yr^{-1}}$ at a $95\%$ confidence level, for a hypothetical hybrid array of radio antennas and scintillators on the full GRANDProto300 layout. These limits are competitive with existing limits in the same energy range, showcasing the potential of this method. This opens a yet uncharted territory in photon searches at UHE, by focusing on higher zenith angles and very inclined showers. 

Future work could enhance discrimination performances by incorporating additional observables, such as the number of activated scintillators, and exploring multivariate analyses or machine learning methods. Applying this method to planned large-scale radio observatories of extensive air showers like GRAND or Auger-Radio could showcase their potential for the search of UHE photons, opening a new window on the acceleration of cosmic rays at UHE.

\appendix
\section{Parameter influence on radio and particle signals}\label{sec:appendixA}

\begin{figure}
    \centering
    \includegraphics[width=\linewidth]{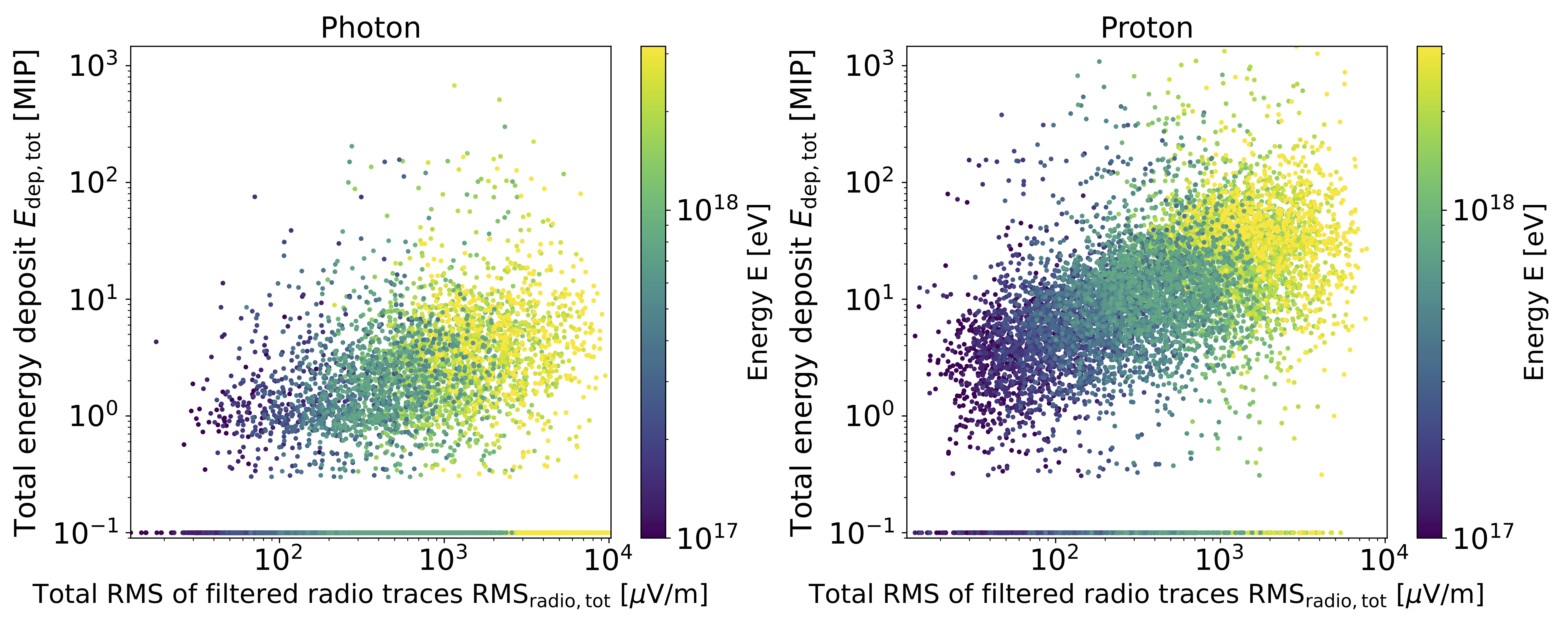}
    \caption{Total energy deposit vs. total RMS of the radio traces of triggered antennas, for photon (\textit{left}) and proton (\textit{right}) events. The primary energy $E$ is color-coded.}
    \label{fig:scatter_energy}
\end{figure}

\begin{figure}
    \centering
    \includegraphics[width=\linewidth]{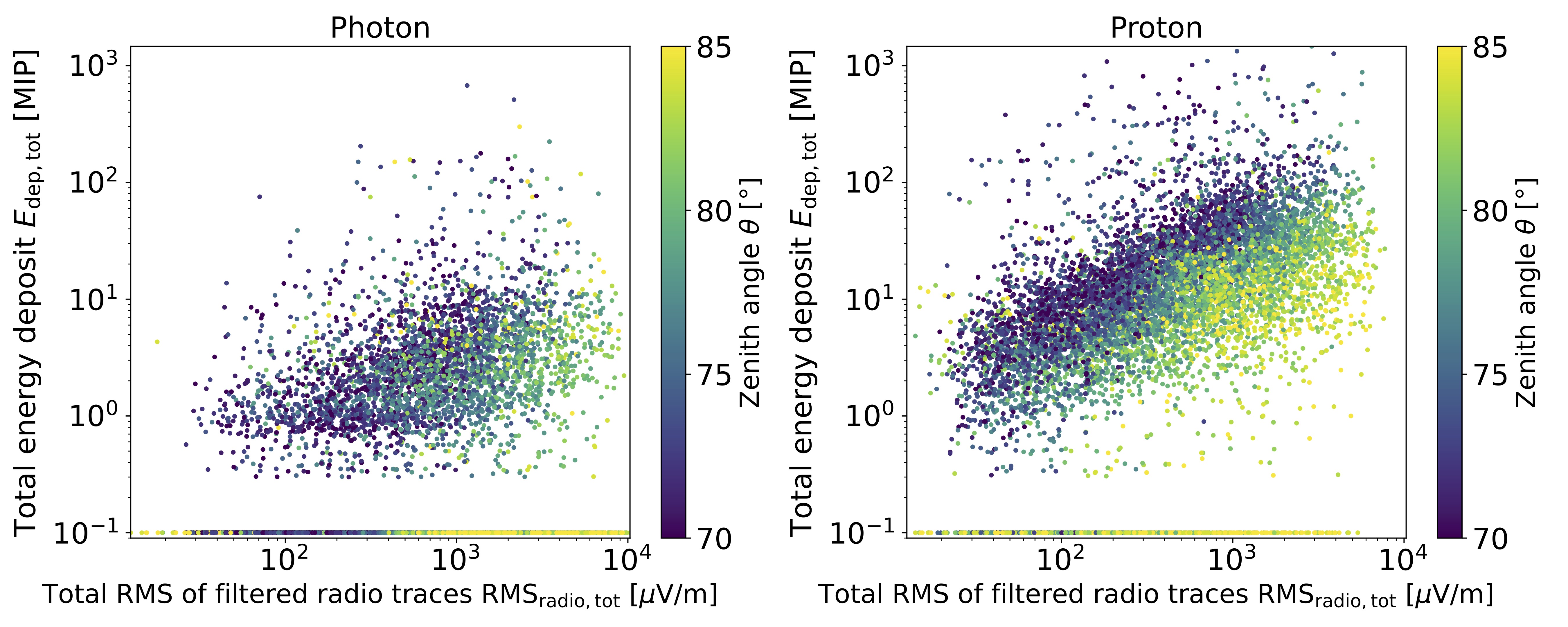}
    \caption{Same as Fig.~\ref{fig:scatter_energy} with the zenith angle $\theta$ color-coded.
    }
    \label{fig:scatter_zenith}
\end{figure}

The distribution of the total energy deposit $E_{\rm dep,tot}$ and the total RMS of the filtered radio traces $\rm RMS_{radio,tot}$ exhibits significant correlations with the energy $E$ and arrival zenith angle $\theta$ of the primary particle. Particles with a lower primary energy tend to have lower $\rm RMS_{radio,tot}$ as well as lower $E_{\rm dep,tot}$, although the total energy deposit varies a lot for showers with the same primary energy, as shown in Figure~\ref{fig:scatter_energy}. Figure~\ref{fig:scatter_zenith} indicates that showers with high zenith angles have a lower total energy deposit, both for proton and photon showers, and that the influence of the primary zenith angle is dominant compared to that of the primary energy. 

\begin{figure}
    \centering
    \includegraphics[width=\linewidth]{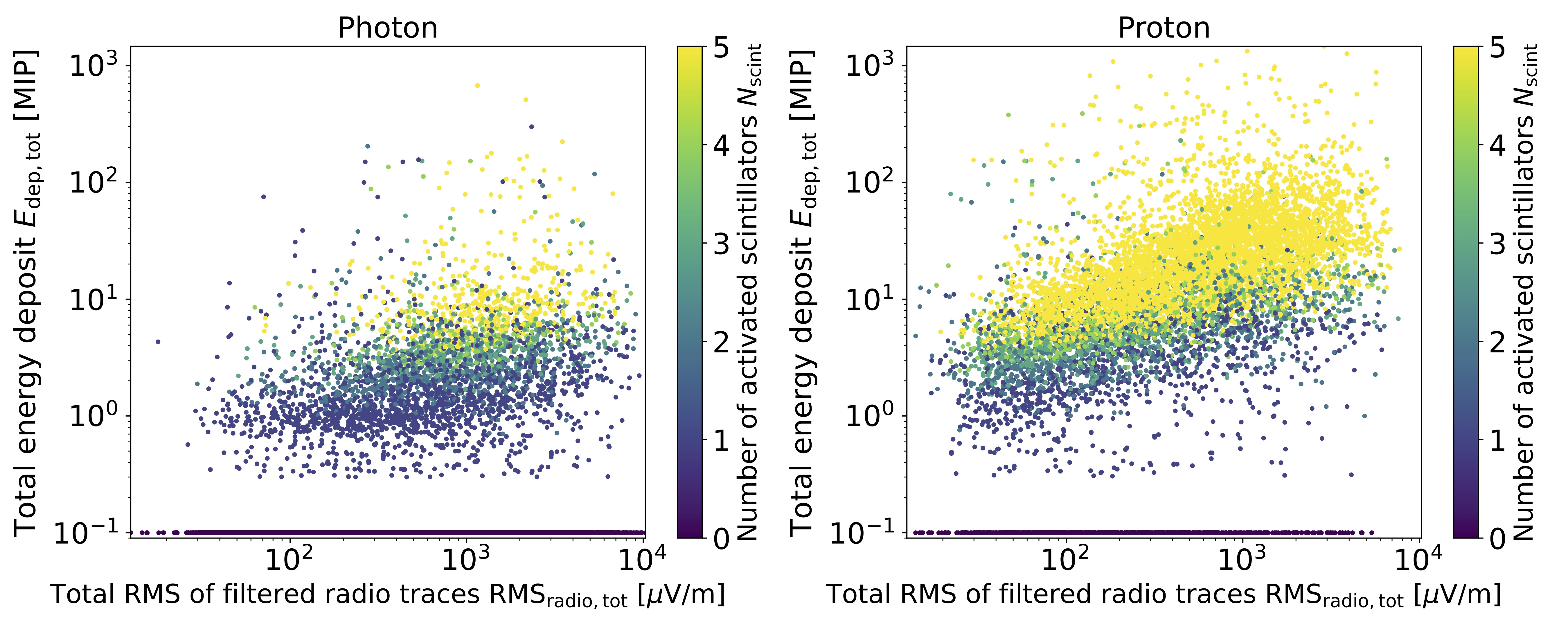}
    \caption{Same as Fig.~\ref{fig:scatter_energy}
    with the color representing the number of activated scintillators $N_{\rm scint}$, with a total energy deposit above the threshold $\epsilon'_{\rm thr} = 0.3 \,\rm MIP$, with a maximum value of $N_{\rm scint} = 5$ for the color scale.}
    \label{fig:scatter_nde}
\end{figure}

Additionally, the number of activated scintillators $N_{\rm scint}$, i.e., scintillators with an energy deposit above the threshold $\epsilon'_{\rm thr} = 0.3\,\rm MIP$, is a clear difference between proton and photon showers, as shown by Fig.~\ref{fig:scatter_nde}. Most proton events have more than 5 activated scintillators, while for most photon events, $N_{\rm scint} \leq 4$. Using $N_{\rm scint}$ as an additional parameter for the classification, for example by performing a Multi-Variate Analysis (MVA), could therefore improve the performances of our method.

\section{Exploration of threshold choice}\label{sec:appendixB}

\begin{figure}
    \centering
    \includegraphics[width=\linewidth]{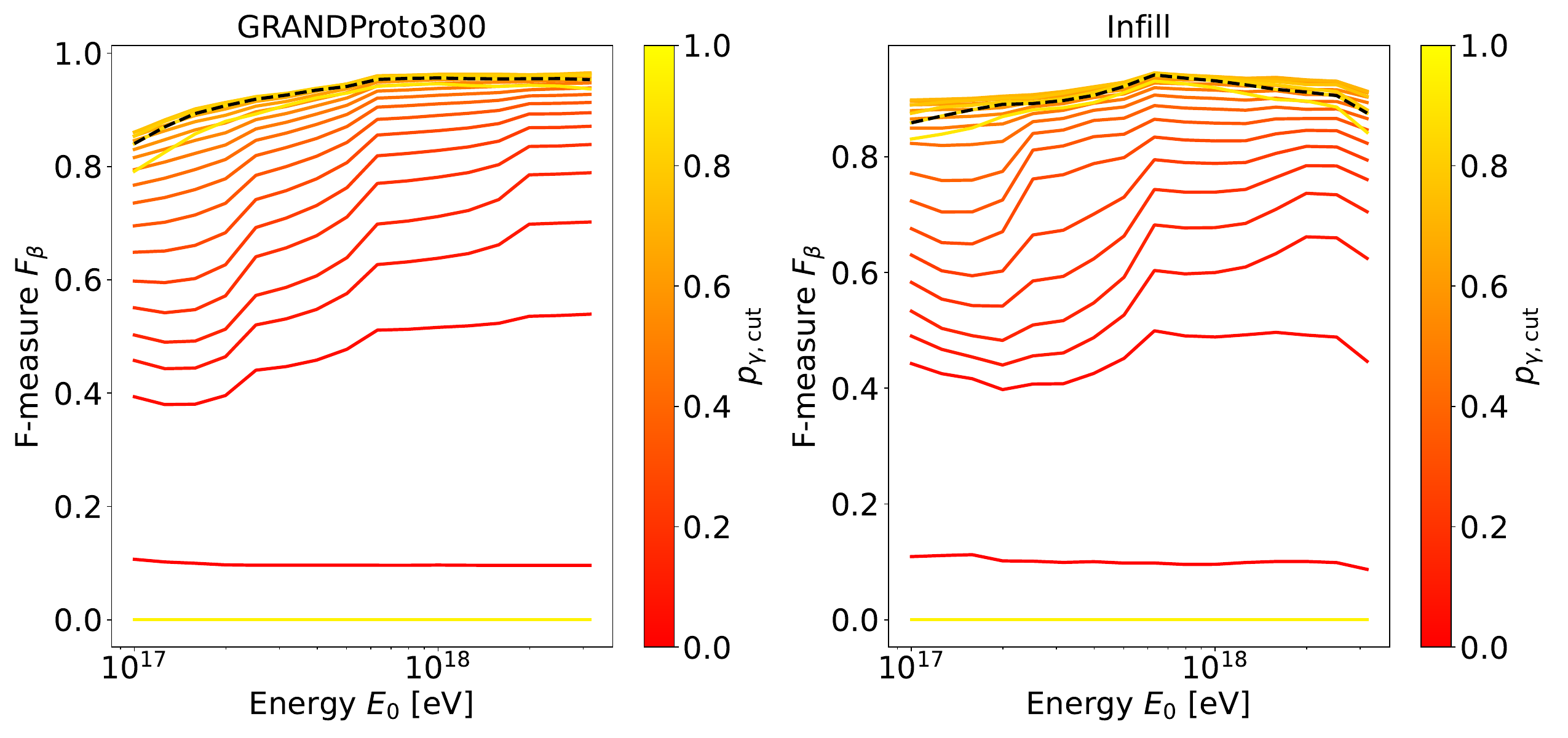}
    \caption{Weighted efficiency and precision estimator $F(\beta=0.25)$-score (see Eq.~\ref{eq:fbeta}) as a function of the primary particle energy $E_0$, with the threshold $p_{\gamma, \rm cut}$ color-coded. The value $p_{\gamma, \rm cut}=0.9$ is identified with a dashed black line.}
    \label{fig:f-measure}
\end{figure}

\begin{figure}
    \centering
    \includegraphics[width=0.6\linewidth]{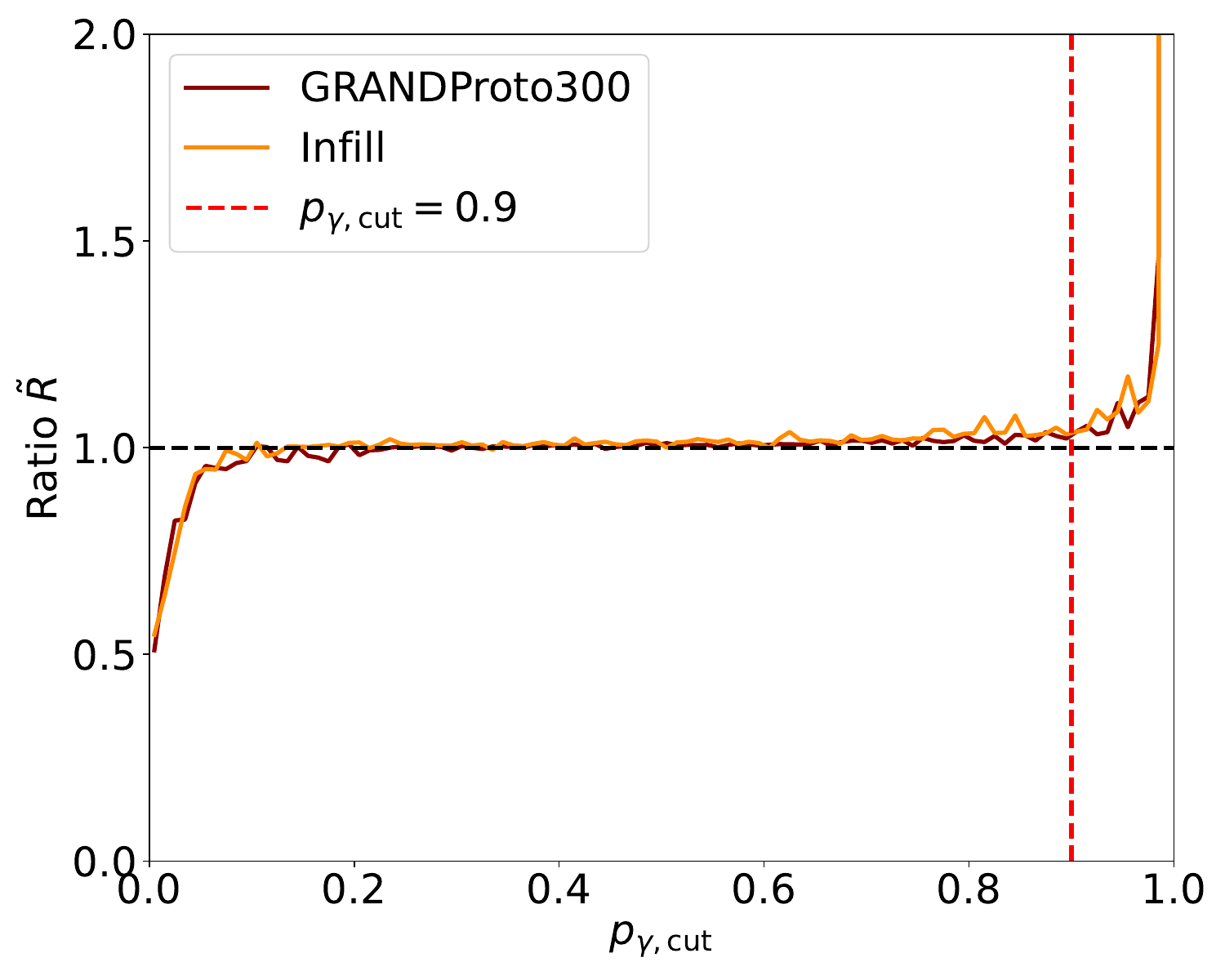}
    \caption{Evolution of the ratio between the efficiency and the contamination, represented by $\tilde{R}(p_{\gamma,\rm cut})$, vs. the photon candidate cut threshold $p_{\gamma,\rm cut}$, for both GRANDProto300 and infill arrays. $\tilde{R}=1$ is shown as a solid black line and the chosen $p_{\gamma,\rm cut}=0.9$ as a dashed red line.}
    \label{fig:ratio}
\end{figure}

The choice of the optimal threshold $p_{\gamma,\rm cut}$ to identify photon events based on the probability for an event to be a photon event $p_{\gamma}$, predicted by the classifier, relies on finding an optimal compromise between a low background contamination and a high enough signal efficiency. The background contamination can be measured using the precision metric, defined as $p = {\rm TP}/({\rm TP+FP})$, with $\rm FP$ the number of proton events misidentified as photon events ("false positives"), and $\rm TP$ the number of photon events correctly classified as photon events ("true positives"). The signal efficiency is measured by the recall, also called sensitivity, $r = {\rm TP}/({\rm TP+FN})$ with $\rm FN$ the number of photon events misidentified as proton events ("false negatives"). A high threshold $p_{\gamma,\rm cut}$ will result in a precision close to 1, but at the cost of a poor signal efficiency, while a low threshold is associated to a poor precision but a signal efficiency close to 1. In order to combine these two measures into one, one can consider the $F(\beta)$-score, defined as the weighted harmonic mean of precision and recall:
\begin{equation}\label{eq:fbeta}
    F(\beta) = \frac{(1+\beta^2)pr}{\beta^2p+r} = \frac{(1+\beta^2)\,\rm TP}{(1+\beta^2)\rm TP+\beta^2FN + FP}\ .
\end{equation}
Choosing a value of $\beta<1$ provides more weight to the precision, while $\beta>1$ focuses on the recall. Because we want to identify photon events in a dominant proton background, we enhance the weight on precision, hence choose $\beta = 0.25$ as an illustrative optimized value.
$\rm TP$, $\rm FP$ and $\rm FN$ are computed as:
\begin{align}
    {\rm TP} &= \frac{1}{c_E(\Gamma)}\int_{E_0}^{\infty}\int_0^{\frac{\pi}{2}}E_\gamma^{-\Gamma}\tau(E_\gamma, \theta) \varepsilon_{\gamma,\textrm{cut}}(E_\gamma,\theta)\cos{\theta} \sin{\theta} \,\textrm{d}\theta \, \textrm{d}E_\gamma \ , \\
    {\rm FN} &= \frac{1}{c_E(\Gamma)}\int_{E_0}^{\infty}\int_0^{\frac{\pi}{2}}E_\gamma^{-\Gamma}\tau(E_\gamma, \theta)(1- \varepsilon_{\gamma,\textrm{cut}}(E_\gamma,\theta))\cos{\theta} \sin{\theta} \,\textrm{d}\theta \, \textrm{d}E_\gamma \ ,\\
    {\rm FP} &= \frac{w_{\rm p}}{c_E(\Gamma_{\rm p})}\int_{E_0}^{\infty}\int_0^{\frac{\pi}{2}}E_{\rm p}^{-\Gamma_{\rm p}}\tau_{\rm p}(E_{\rm p}, \theta) \xi_{\textrm{p,cut}}(E_{\rm p},\theta)\cos{\theta} \sin{\theta} \,\textrm{d}\theta \, \textrm{d}E_{\rm p}\ ,
\end{align}
where $c_E$ is the normalization constant defined in Eq.~\ref{eq:ce} and $w_{\rm p} = 10$ is a weighting factor assigned to proton events to account for the difference in fluxes between proton and photon events. We do not add the factors accounting for the area of the detector and the time of operation since they will cancel out when computing ratios. $F(\beta=0.25)$ is shown as a function of energy for various $p_{\gamma,\rm cut}$ thresholds in Fig.~\ref{fig:f-measure}. For the full GRANDProto300 layout, thresholds between $0.6$ and $0.9$ reach comparable scores. The infill reaches better performances for slightly lower thresholds.
Another metric we can use is $R(p_{\gamma,\rm cut},E_0) = {\bar{\varepsilon}_{\gamma, \rm cut}(p_{\gamma,\rm cut},E_0)}/{\bar{\xi}_{\rm p,cut}(p_{\gamma,\rm cut},E_0)}$, the ratio of efficiency and contamination and the way it varies when increasing the $p_{\gamma,\rm cut}$ threshold. In particular we look at:
\begin{eqnarray}
    \tilde{R}(p_{\gamma,\rm cut}) &=& \max_{E_0} \frac{R(p_{\gamma,\rm cut} + \Delta p_{\gamma,\rm cut}, E_0)}{R(p_{\gamma,\rm cut},E_0)} \nonumber\\
    &=& \max_{E_0}\left[{\frac{\bar{\varepsilon}_{\gamma, \rm cut}(p_{\gamma,\rm cut} + \Delta p_{\gamma,\rm cut}, E_0)}{\bar{\varepsilon}_{\gamma, \rm cut}(p_{\gamma,\rm cut}, E_0)}}{\frac{\bar{\xi}_{\rm p,cut}(p_{\gamma,\rm cut}, E_0)}{\bar{\xi}_{\rm p,cut}(p_{\gamma,\rm cut} + \Delta p_{\gamma,\rm cut}, E_0)}}\right] \ ,
\end{eqnarray}
with $\bar{\varepsilon}_{\gamma, \rm cut}$ and $\bar{\xi}_{\rm p,cut}$ as defined in Eqs.~\ref{eq:effi} and \ref{eq:xi}.
$\tilde{R}(p_{\gamma,\rm cut})$ represents how the loss in signal efficiency is compensated by the improvement of the background contamination, when one increases the threshold $p_{\gamma,\rm cut}$ to $p_{\gamma,\rm cut}+\Delta p_{\gamma,\rm cut}$, and it is shown in Figure~\ref{fig:ratio}, for a value $\Delta p_{\gamma,\rm cut}=0.01$. While $\tilde{R} < 1$, increasing the threshold $p_{\gamma,\rm cut}$ allows to improve the background contamination without losing too much in signal efficiency. When $\tilde{R} \gg 1$, increasing the threshold does not provide further improvement.
Therefore, we find our optimal threshold as the threshold above which $\tilde{R}$ rises significantly above 1. According to Fig.~\ref{fig:ratio}, $p_{\gamma,\rm cut}=0.9$ is a typical optimal value for the GRANDProto300 array, although lower cut values might be more adapted for the infill. We keep $p_{\gamma,\rm cut}=0.9$ for the photon candidate cut, so that our upper limits are conservative.

\section{Impact of reconstruction resolution on flux limits}\label{sec:appendixC}

\begin{figure}
    \centering
    \includegraphics[width=0.8\linewidth]{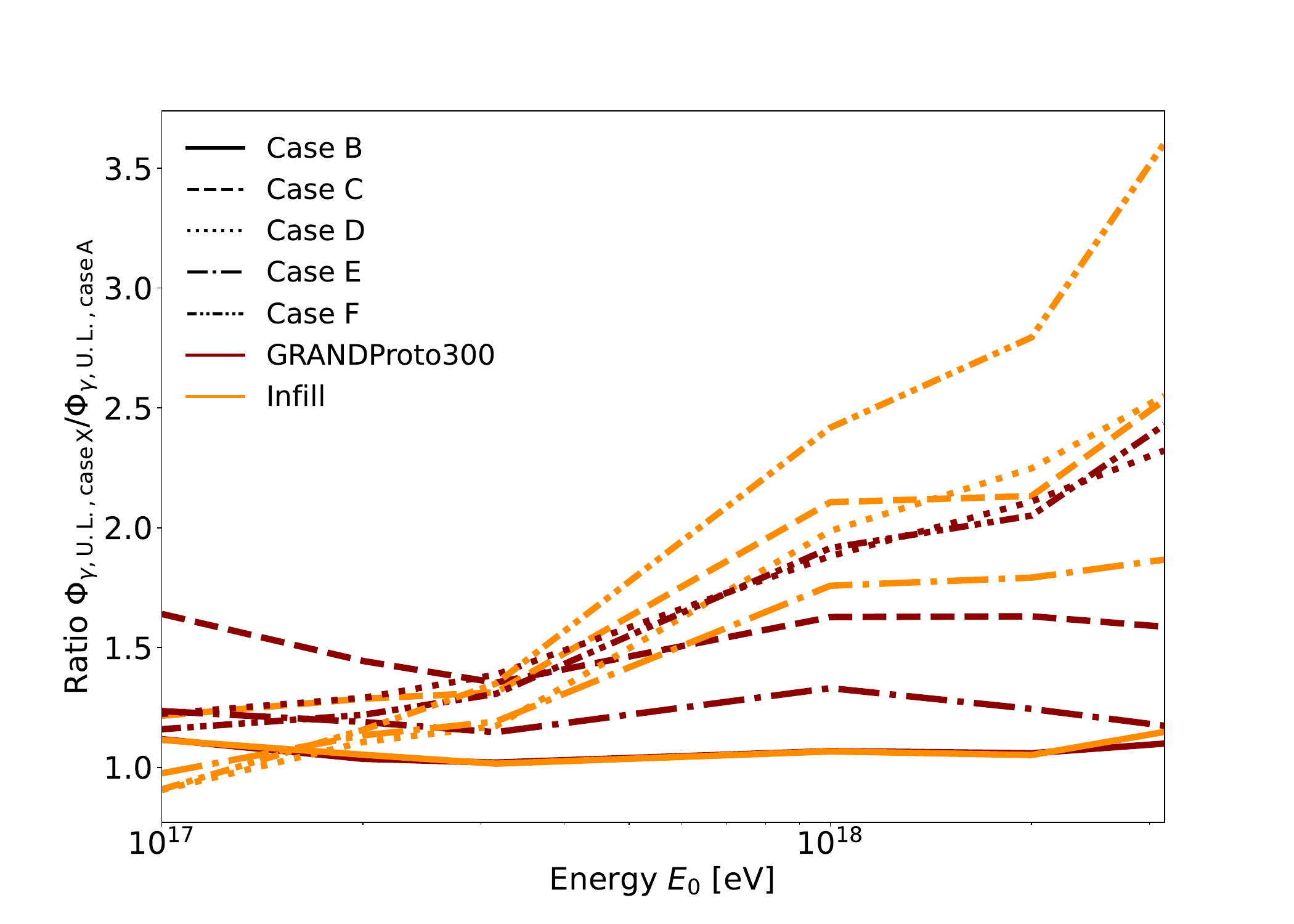}
    \caption{Ratio $\Phi_{\gamma,\rm U.L., case\,X}/\Phi_{\gamma,\rm U.L., case\,A}$ of the upper limits obtained for the various cases studied X\,$=$\,B--F (see Table~\ref{tab:cases}) relative to reference case A. Results are shown for both the GRANDProto300 and infill arrays. The parameters for each case are detailed in Table~\ref{tab:cases}.}
    \label{fig:flux_lim_ratios}
\end{figure}

As described in Section~\ref{sec:z_only}, we investigate the impact of the reconstruction resolution of the primary particle energy $E$ and zenith angle $\theta$ by varying the energy bin width, $\Delta\log_{10}(E/\rm eV)$, and the zenith bin width, $\Delta\theta$. To assess the influence of energy, we test configurations with and without energy binning; to assess the influence of zenith resolution, we test both broad ($\Delta\theta = 4^\circ$) and thin ($\Delta\theta = 1^\circ$) zenith bins. Two sample weighting functions are tested: $w(E) \propto E^{-2}$, reflecting the expected photon flux spectrum, and uniform weights $w(E) = 1$, which prioritize high-energy events to enhance classification performance. The configurations for each case are summarized in Table~\ref{tab:cases}.
To ensure comparability across all cases, the photon probability threshold $p_{\gamma, \rm cut}$ is adjusted so that the average contamination remains around $0.1\%$—the reference contamination for case A with $p_{\gamma, \rm cut}=0.9$ being $0.12\%$. This results in $p_{\gamma, \rm cut} = 0.9$ for case B, $0.95$ for cases C to F. Figure~\ref{fig:flux_lim_ratios} presents the ratio of the computed upper limits for each case relative to the reference case A, for both the full GRANDProto300 and the infill arrays.
For case C (broad zenith bins with sample weighting), the upper limits increase by up to a factor of $~1.7$ for the infill array and $~2.5$ for GRANDProto300. When using thin zenith bins with sample weights (case E), the upper limits increase on average by $45\%$ for GRANDProto300, and $20\%$ for the infill array, demonstrating the method’s robustness even without energy information in the analysis. When removing the energy binning, using uniform sample weights ($w(E) = 1$) instead of flux-proportional weights ($w(E) \propto E^{-2}$) to compute the classification boundary results in comparable upper limits at lower energies, because of the adjusted photon probability threshold $p_{\gamma, \rm cut}$. However, the associated increase in $p_{\gamma, \rm cut}$ results in a lower signal efficiency at high energies, hence higher upper limits. This approach makes the obtained upper limits ratios conservative, especially at high energies. In general, using flux-proportional weights ($w(E) \propto E^{-2}$) for the classification yields the best performances.

\begin{table}[]
    \centering
    \resizebox{\textwidth}{!}{
    \begin{tabular}{lccc}
    \toprule
        \textbf{Case} & \begin{tabular}{@{}c@{}} \textbf{Energy bins} \\ $\Delta \log_{10}(E/\rm eV)$\end{tabular} & \begin{tabular}{@{}c@{}} \textbf{Zenith bins} \\ $\Delta\theta$ (°)\end{tabular} & \begin{tabular}{@{}c@{}} \textbf{Sample weighting} \\ $w(E)$\end{tabular} \\
        \midrule
         A (Reference) & 0.4 & 4  & $E^{-2}$\\
         B (No weights) & 0.4 & 4 & 1 \\
         C (No energy bins) & — & 4 & $E^{-2}$ \\
         D (No energy bins, no weights) & — & 4 & 1 \\
         E (No energy bins, thin zenith bins) & — & 1 & $E^{-2}$ \\
         F (No energy bins, thin zenith bins, no weights) & — & 1 & 1 \\
         \bottomrule
    \end{tabular}}
    \caption{Summary of the cases considered to evaluate the impact of energy and zenith angle reconstruction resolution on the classification performance. Case A serves as the reference configuration, detailed in Sec.~\ref{sec:discr}. For each subsequent case, modifications relative to the reference for energy bin width, $\Delta\log_{10}(E/\rm eV)$, zenith angle bin width, $\Delta\theta$, and the sample weighting function, $w(E)$ are indicated.
    The sample weighting $w(E)$ is either proportional to $E^{-2}$, reflecting the expected photon flux spectrum, or uniform, $w(E) = 1$, for a better classification at higher energies.}
    \label{tab:cases}
\end{table}

\acknowledgments
The authors thank Rafael Alves Batista, Keitaro Fujita, Sei Kato, Kazumasa Kawata, Jelena Koehler, Hitoshi Oshima, Tanguy Pierog, the GRAND Collaboration, and particularly the GRAND-Paris team, and the ICRR TA team, for fruitful discussions, as well as the TA collaboration for providing detector response simulation codes. P.\,M. acknowledges support from the Erasmus+ program, the MESRI AMX grant. K.\,K. thanks ICRR and ILANCE for the hospitality and the support. K.\,K. acknowledges support from the Fulbright-France program, the CNRS Programme Blanc MITI ("GRAND" 2023.1 268448; France), the CNRS Programme AMORCE ("GRAND" 258540; France). K.T. and T.S. acknowledge support from the CNRS-University of Tokyo PhD Joint Program.

\bibliographystyle{JHEP}
\bibliography{JCAP}

\end{document}